# Automated Reasoning in Modal and Description Logics via SAT Encoding: the Case Study of $K_m/\mathcal{ALC}$-Satisfiability

**Roberto Sebastiani**                    ROBERTO.SEBASTIANI@DISI.UNITN.IT

**Michele Vescovi**                       MICHELE.VESCOVI@DISI.UNITN.IT

*DISI, Università di Trento*
*Via Sommarive 14, I-38123, Povo, Trento, Italy*

## Abstract

In the last two decades, modal and description logics have been applied to numerous areas of computer science, including knowledge representation, formal verification, database theory, distributed computing and, more recently, semantic web and ontologies. For this reason, the problem of automated reasoning in modal and description logics has been thoroughly investigated. In particular, many approaches have been proposed for efficiently handling the satisfiability of the core normal modal logic $K_m$, and of its notational variant, the description logic $\mathcal{ALC}$. Although simple in structure, $K_m/\mathcal{ALC}$ is computationally very hard to reason on, its satisfiability being PSPACE-complete.

In this paper we start exploring the idea of performing automated reasoning tasks in modal and description logics by encoding them into SAT, so that to be handled by state-of-the-art SAT tools; as with most previous approaches, we begin our investigation from the satisfiability in $K_m$. We propose an efficient encoding, and we test it on an extensive set of benchmarks, comparing the approach with the main state-of-the-art tools available. Although the encoding is necessarily worst-case exponential, from our experiments we notice that, in practice, this approach can handle most or all the problems which are at the reach of the other approaches, with performances which are comparable with, or even better than, those of the current state-of-the-art tools.

## 1. Motivations and Goals

In the last two decades, modal and description logics have provided an essential framework for many applications in numerous areas of computer science, including artificial intelligence, formal verification, database theory, distributed computing and, more recently, semantic web and ontologies. For this reason, the problem of automated reasoning in modal and description logics has been thoroughly investigated (e.g., Fitting, 1983; Ladner, 1977; Baader & Hollunder, 1991; Halpern & Moses, 1992; Baader, Franconi, Hollunder, Nebel, & Profitlich, 1994; Massacci, 2000). In particular, the research in modal and description logics has followed two parallel routes until the seminal work of Schild (1991), which proved that the core modal logic $K_m$ and the core description logic $\mathcal{ALC}$ are one a notational variant of the other. Since then, analogous results have been produced for a bunch of other logics, so that, nowadays the two research lines have mostly merged into one research flow.

Many approaches have been proposed for efficiently reasoning in modal and description logics, starting from the problem of checking the satisfiability in the core normal modal logic $K_m$ and in its notational variant, the description logic $\mathcal{ALC}$ (hereafter simply "$K_m$"). We classify them as follows.

- The "classic" *tableau-based* approach (Fitting, 1983; Baader & Hollunder, 1991; Massacci, 2000) is based on the construction of propositional tableau branches, which are recursively expanded on demand by generating successor nodes in a candidate Kripke model. KRIS (Baader & Hollunder, 1991; Baader et al., 1994), CRACK (Franconi, 1998), LWB (Balsiger, Heuerding, & Schwendimann, 1998) were among the main representative tools of this approach.

- The *DPLL-based* approach (Giunchiglia & Sebastiani, 1996, 2000) differs from the previous one mostly in the fact that a Davis-Putnam-Logemann-Loveland (DPLL) procedure, which treats the modal subformulas as propositions, is used instead of the classic propositional tableaux procedure at each nesting level of the modal operators. KSAT (Giunchiglia & Sebastiani, 1996), ESAT (Giunchiglia, Giunchiglia, & Tacchella, 2002) and *SAT (Tacchella, 1999), are the representative tools of this approach.

These two approaches merged into the "modern" tableaux-based approach, which has been extended to work with more expressive description logics and to provide more sophisticate reasoning functions. Among the tools employing this approach, we recall FACT/FACT++ and DLP (Horrocks & Patel-Schneider, 1999), and RACER (Haarslev & Moeller, 2001). [1]

- In the *translational* approach (Hustadt & Schmidt, 1999; Areces, Gennari, Heguiabehere, & de Rijke, 2000) the modal formula is encoded into first-order logic (FOL), and the encoded formula can be decided efficiently by a FOL theorem prover (Areces et al., 2000). MSPASS (Hustadt, Schmidt, & Weidenbach, 1999) is the most representative tool of this approach.

- The *CSP-based approach* (Brand, Gennari, & de Rijke, 2003) differs from the tableaux-based and DPLL-based ones mostly in the fact that a CSP (Constraint Satisfaction Problem) engine is used instead of tableaux/DPLL. KCSP is the only representative tool of this approach.

- In the *Inverse-method* approach (Voronkov, 1999, 2001), a search procedure is based on the "inverted" version of a sequent calculus (which can be seen as a modalized version of propositional resolution). КЯ (Voronkov, 1999) is the only representative tool of this approach.

- In the *Automata-theoretic* approach, (a symbolic representation based on BDDs – Binary Decision Diagrams – of) a tree automaton accepting all the tree models of the input formula is implicitly built and checked for emptiness (Pan, Sattler, & Vardi, 2002; Pan & Vardi, 2003). KBDD (Pan & Vardi, 2003) is the only representative tool of this approach.

---

[1]. Notice that there is not an universal agreement on the terminology "tableaux-based" and "DPLL-based". E.g., tools like FACT, DLP, and RACER are most often called "tableau-based", although they use a DPLL-like algorithm instead of propositional tableaux for handling the propositional component of reasoning (Horrocks, 1998; Patel-Schneider, 1998; Horrocks & Patel-Schneider, 1999; Haarslev & Moeller, 2001).

- Pan and Vardi (2003) presented also an encoding of K-satisfiability into QBF-satisfiability (which is PSPACE-complete too), combined with the use of a state-of-the-art QBF (Quantified Boolean Formula) solver. We call this approach *QBF-encoding* approach.

To the best of our knowledge, the last four approaches so far are restricted to the satisfiability in $K_m$ only, whilst the translational approach has been applied to numerous modal and description logics (e.g. traditional modal logics like $T_m$ and $S4_m$, and dynamic modal logics) and to the relational calculus.

A significant amount of benchmarks formulas have been produced for testing the effectiveness of the different techniques (Halpern & Moses, 1992; Giunchiglia, Roveri, & Sebastiani, 1996; Heuerding & Schwendimann, 1996; Horrocks, Patel-Schneider, & Sebastiani, 2000; Massacci, 1999; Patel-Schneider & Sebastiani, 2001, 2003).

In the last two decades we have also witnessed an impressive advance in the efficiency of propositional satisfiability techniques (SAT), which has brought large and previously-intractable problems at the reach of state-of-the-art SAT solvers. Most of the success of SAT technologies is motivated by the impressive efficiency reached by current implementations of the DPLL procedure, (Davis & Putnam, 1960; Davis, Longemann, & Loveland, 1962), in its most-modern variants (Silva & Sakallah, 1996; Moskewicz, Madigan, Zhao, Zhang, & Malik, 2001; Eén & Sörensson, 2004). Current implementations can handle formulas in the order of $10^7$ variables and clauses.

As a consequence, many hard real-world problems have been successfully solved by encoding into SAT (including, e.g., circuit verification and synthesis, scheduling, planning, model checking, automatic test pattern generation , cryptanalysis, gene mapping). Effective encodings into SAT have been proposed also for the satisfiability problems in quantifier-free FOL theories which are of interest for formal verification (Strichman, Seshia, & Bryant, 2002; Seshia, Lahiri, & Bryant, 2003; Strichman, 2002). Notably, successful SAT encodings include also PSPACE-complete problems, like planning (Kautz, McAllester, & Selman, 1996) and model checking (Biere, Cimatti, Clarke, & Zhu, 1999).

In this paper we start exploring the idea of performing automated reasoning tasks in modal and description logics by encoding them into SAT, so that to be handled by state-of-the-art SAT tools; as with most previous approaches, we begin our investigation from the satisfiability in $K_m$.

In theory, the task may look hopeless because of worst-case complexity issues: in fact, with few exceptions, the satisfiability problem in most modal and description logics is not in NP, typically being PSPACE-complete or even harder —PSPACE-complete for $K_m$ (Ladner, 1977; Halpern & Moses, 1992)— so that the encoding is in worst-case non polynomial. [2]

In practice, however, a few considerations allow for not discarding that this approach may be competitive with the state-of-the-art approaches. First, the non-polynomial bounds above are *worst-case* bounds, and formulas may have different behaviors from that of the pathological formulas which can be found in textbooks. (E.g., notice that the exponentiality is based on the hypothesis of unboundedness of some parameter like the modal depth; Halpern & Moses, 1992; Halpern, 1995.) Second, some tricks in the encoding may allow for reducing the size of the encoded formula significantly. Third, as the amount of RAM memory in current computers is in the order of the GBytes and current SAT solvers can successfully handle huge formulas, the encoding of many modal formulas (at least of those which are not too hard to solve also for the competitors) may be at the reach of a SAT solver. Finally, even for PSPACE-complete logics like $K_m$, also other state-of-the-art approaches are not guaranteed to use polynomial memory.

---

2. We implicitly make the assumption NP $\neq$ PSPACE.

In this paper we show that, at least for the satisfiability $K_m$, by exploiting some smart optimizations in the encoding the SAT-encoding approach becomes competitive in practice with previous approaches. To this extent, the contributions of this paper are manyfold.

- We propose a basic encoding of $K_m$ formulas into purely-propositional ones, and prove that the encoding is satisfiability-preserving.

- We describe some optimizations of the encoding, both in form of preprocessing and of on-the-fly simplification. These techniques allow for significant (and in some cases dramatic) reductions in the size of the resulting Boolean formulas, and in performances of the SAT solver thereafter.

- We perform a very extensive empirical comparison against the main state-of-the-art tools available. We show that, despite the NP-vs.-PSPACE issue, this approach can handle most or all the problems which are at the reach of the other approaches, with performances which are comparable with, and sometimes even better than, those of the current state-of-the-art tools. In our perspective, this is the most surprising contribution of the paper.

- As a byproduct of our work, we obtain an empirical evaluation of current tools for $K_m$-satisfiability available, which is very extensive in terms of both amount and variety of benchmarks and of number and representativeness of the tools evaluated. We are not aware of any other such evaluation in the recent literature.

We also stress the fact that with our approach the encoder can be interfaced with every SAT solver in a plug-and-play manner, so that to benefit for free of every improvement in the technology of SAT solvers which has been or will be made available.

**Content.** The paper is structured as follows. In Section 2 we provide the necessary background notions on modal logics and SAT. In Section 3 we describe the basic encoding from $K_m$ to SAT. In Section 4 we describe and discuss the main optimizations, and provide many examples. In Section 5 we present the empirical evaluation, and discuss the results. In Section 6 we present some related work and current research trends. In Section 7 we conclude, and describe some possible future evolutions.

A six-page preliminary version of this paper, containing some of the basic ideas presented here, was presented at SAT'06 conference (Sebastiani & Vescovi, 2006). For the readers' convenience, an online appendix is provided, containing all plots of Section 5 in full size. Moreover, in order to make the results reproducible, the encoder, the benchmarks and the random generators with the seeds used are also available in the online appendix.

## 2. Background

In this section we provide the necessary background on the modal logic $K_m$ (Section 2.1) and on SAT and the DPLL procedure (Section 2.2).

### 2.1 The Modal Logic $K_m$

We recall some basic definitions and properties of $K_m$. Given a non-empty set of primitive propositions $\mathcal{A} = \{A_1, A_2, \ldots\}$, a set of $m$ modal operators $\mathcal{B} = \{\Box_1, \ldots, \Box_m\}$, and the constants "*True*" and "*False*" (that we denote respectively with "$\top$" and "$\bot$") the language of $K_m$ is the least set of formulas containing $\mathcal{A}$, closed under the set of propositional connectives $\{\neg, \wedge, \vee, \rightarrow, \leftrightarrow\}$ and the set of modal operators in $\mathcal{B} \cup \{\Diamond_1, \ldots, \Diamond_m\}$. Notationally, we use the Greek letters $\alpha, \beta, \varphi, \psi, \nu, \pi$ to denote formulas in the language of $K_m$ ($K_m$-formulas hereafter). Notice that we can consider $\{\neg, \wedge\}$ together with $\mathcal{B}$ as the group of the primitive connectives/operators, defining the remaining in the standard way, that is: "$\Diamond_r\varphi$" for "$\neg\Box_r\neg\varphi$", "$\varphi_1 \vee \varphi_2$" for "$\neg(\neg\varphi_1 \wedge \neg\varphi_2)$", "$\varphi_1 \rightarrow \varphi_2$" for "$\neg(\varphi_1 \wedge \neg\varphi_2)$", "$\varphi_1 \leftrightarrow \varphi_2$" for "$\neg(\varphi_1 \wedge \neg\varphi_2) \wedge \neg(\varphi_2 \wedge \neg\varphi_1)$". (Hereafter formulas like $\neg\neg\psi$ are implicitly assumed to be simplified into $\psi$, so that, if $\psi$ is $\neg\phi$, then by "$\neg\psi$" we mean "$\phi$".) Notationally, we often write "$(\bigwedge_i l_i) \rightarrow \bigvee_j l_j$" for the clause "$\bigvee_j \neg l_i \vee \bigvee_j l_j$", and "$(\bigwedge_i l_i) \rightarrow (\bigwedge_j l_j)$" for the conjunction of clauses "$\bigwedge_j(\bigvee_i \neg l_i \vee l_j)$". Further, we often write $\Box_r$ or $\Diamond_r$ meaning one specific/generic modal operator, where it is assumed that $r = 1, \ldots, m$; and we denote by $\Box_r^i$ the nested application of the $\Box_r$ operator $i$ times: $\Box_r^0\psi := \psi$ and $\Box_r^{i+1}\psi := \Box_r\Box_r^i\psi$. We call *depth* of $\varphi$, written *depth($\varphi$)*, the maximum number of nested modal operators in $\varphi$. We call a *propositional atom* every primitive proposition in $\mathcal{A}$, and a *propositional literal* every propositional atom (*positive literal*) or its negation (*negative literal*). We call a *modal atom* every formula which is either in the form $\Box_r\varphi$ or in the form $\Diamond_r\varphi$.

In order to make our presentation more uniform, and to avoid considering the polarity of subformulas, we adopt the traditional representation of $K_m$-formulas (introduced, as far as we know, by Fitting, 1983 and widely used in literature, e.g. Fitting, 1983; Massacci, 2000; Donini & Massacci, 2000) from the following table:

| $\alpha$ | $\alpha_1$ | $\alpha_2$ | $\beta$ | $\beta_1$ | $\beta_2$ | $\pi^r$ | $\pi_0^r$ | $\nu^r$ | $\nu_0^r$ |
|---|---|---|---|---|---|---|---|---|---|
| $(\varphi_1 \wedge \varphi_2)$ | $\varphi_1$ | $\varphi_2$ | $(\varphi_1 \vee \varphi_2)$ | $\varphi_1$ | $\varphi_2$ | $\Diamond_r\varphi_1$ | $\varphi_1$ | $\Box_r\varphi_1$ | $\varphi_1$ |
| $\neg(\varphi_1 \vee \varphi_2)$ | $\neg\varphi_1$ | $\neg\varphi_2$ | $\neg(\varphi_1 \wedge \varphi_2)$ | $\neg\varphi_1$ | $\neg\varphi_2$ | $\neg\Box_r\varphi_1$ | $\neg\varphi_1$ | $\neg\Diamond_r\varphi_1$ | $\neg\varphi_1$ |
| $\neg(\varphi_1 \rightarrow \varphi_2)$ | $\varphi_1$ | $\neg\varphi_2$ | $(\varphi_1 \rightarrow \varphi_2)$ | $\neg\varphi_1$ | $\varphi_2$ | | | | |

in which non-literal $K_m$-formulas are grouped into four categories: $\alpha$'s (conjunctive), $\beta$'s (disjunctive), $\pi$'s (existential), $\nu$'s (universal). Importantly, all such formulas occur in the main formula with positive polarity only. This allows for disregarding the issue of polarity of subformulas.

The semantic of modal logics is given by means of Kripke structures. A *Kripke structure* for $K_m$ is a tuple $\mathcal{M} = \langle \mathcal{U}, \mathcal{L}, \mathcal{R}_1, \ldots, \mathcal{R}_m \rangle$, where $\mathcal{U}$ is a set of states, $\mathcal{L}$ is a function $\mathcal{L} : \mathcal{A} \times \mathcal{U} \longmapsto \{True, False\}$, and each $\mathcal{R}_r$ is a binary relation on the states of $\mathcal{U}$. With an abuse of notation we write "$u \in \mathcal{M}$" instead of "$u \in \mathcal{U}$". We call a *situation* any pair $\mathcal{M}, u$, $\mathcal{M}$ being a Kripke structure and $u \in \mathcal{M}$. The binary relation $\models$ between a modal formula

$\varphi$ and a situation $\mathcal{M}, u$ is defined as follows:

$$\mathcal{M}, u \models \top;$$
$$\mathcal{M}, u \not\models \bot;$$
$$\mathcal{M}, u \models A_i, \ A_i \in \mathcal{A} \quad \Longleftrightarrow \quad \mathcal{L}(A_i, u) = True;$$
$$\mathcal{M}, u \models \neg A_i, \ A_i \in \mathcal{A} \quad \Longleftrightarrow \quad \mathcal{L}(A_i, u) = False;$$
$$\mathcal{M}, u \models \alpha \quad \Longleftrightarrow \quad \mathcal{M}, u \models \alpha_1 \ and \ \mathcal{M}, u \models \alpha_2;$$
$$\mathcal{M}, u \models \beta \quad \Longleftrightarrow \quad \mathcal{M}, u \models \beta_1 \ or \ \mathcal{M}, u \models \beta_2;$$
$$\mathcal{M}, u \models \pi^r \quad \Longleftrightarrow \quad \mathcal{M}, w \models \pi_0^r \text{ for some } w \in \mathcal{U} \text{ s.t. } \mathcal{R}_r(u, w) \text{ holds in } \mathcal{M};$$
$$\mathcal{M}, u \models \nu^r \quad \Longleftrightarrow \quad \mathcal{M}, w \models \nu_0^r \text{ for every } w \in \mathcal{U} \text{ s.t. } \mathcal{R}_r(u, w) \text{ holds in } \mathcal{M}.$$

"$\mathcal{M}, u \models \varphi$" should be read as "$\mathcal{M}, u$ *satisfy* $\varphi$ in $K_m$" (alternatively, "$\mathcal{M}, u$ $K_m$-satisfies $\varphi$"). We say that a $K_m$-formula $\varphi$ is satisfiable in $K_m$ ($K_m$-satisfiable henceforth) if and only if there exist $\mathcal{M}$ and $u \in \mathcal{M}$ s.t. $\mathcal{M}, u \models \varphi$. (When this causes no ambiguity, we sometimes drop the prefix "$K_m$-".) We say that $w$ is a *successor* of $u$ through $\mathcal{R}_r$ iff $\mathcal{R}_r(u, w)$ holds in $\mathcal{M}$.

The problem of determining the $K_m$-satisfiability of a $K_m$-formula $\varphi$ is decidable and PSPACE-complete (Ladner, 1977; Halpern & Moses, 1992), even restricting the language to a single Boolean atom (i.e., $\mathcal{A} = \{A_1\}$; Halpern, 1995); if we impose a bound on the modal depth of the $K_m$-formulas, the problem reduces to NP-complete (Halpern, 1995). For a more detailed description on $K_m$— including, e.g., axiomatic characterization, decidability and complexity results — we refer the reader to the works of Halpern and Moses (1992), and Halpern (1995).

A $K_m$-formula is said to be in *Negative Normal Form (NNF)* if it is written in terms of the symbols $\Box_r$, $\Diamond_r$, $\wedge$, $\vee$ and propositional literals $A_i$, $\neg A_i$ (i.e., if all negations occur only before propositional atoms in $\mathcal{A}$). Every $K_m$-formula $\varphi$ can be converted into an equivalent one $NNF(\varphi)$ by recursively applying the rewriting rules: $\neg \Box_r \varphi \Longrightarrow \Diamond_r \neg \varphi$, $\neg \Diamond_r \varphi \Longrightarrow \Box_r \neg \varphi$, $\neg(\varphi_1 \wedge \varphi_2) \Longrightarrow (\neg \varphi_1 \vee \neg \varphi_2)$, $\neg(\varphi_1 \vee \varphi_2) \Longrightarrow (\neg \varphi_1 \wedge \neg \varphi_2)$, $\neg \neg \varphi \Longrightarrow \varphi$.

A $K_m$-formula is said to be in *Box Normal Form (BNF)* (Pan et al., 2002; Pan & Vardi, 2003) if it is written in terms of the symbols $\Box_r$, $\neg \Box_r$, $\wedge$, $\vee$, and propositional literals $A_i$, $\neg A_i$ (i.e., if no diamonds are there, and all negations occur only before boxes or before propositional atoms in $\mathcal{A}$). Every $K_m$-formula $\varphi$ can be converted into an equivalent one $BNF(\varphi)$ by recursively applying the rewriting rules: $\Diamond_r \varphi \Longrightarrow \neg \Box_r \neg \varphi$, $\neg(\varphi_1 \wedge \varphi_2) \Longrightarrow (\neg \varphi_1 \vee \neg \varphi_2)$, $\neg(\varphi_1 \vee \varphi_2) \Longrightarrow (\neg \varphi_1 \wedge \neg \varphi_2)$, $\neg \neg \varphi \Longrightarrow \varphi$.

## 2.2 Propositional Satisfiability with the DPLL Algorithm

Most state-of-the-art SAT procedures are evolutions of the DPLL procedure (Davis & Putnam, 1960; Davis et al., 1962). A high-level schema of a modern DPLL engine, adapted from the one presented by Zhang and Malik (2002), is reported in Figure 1. The Boolean formula $\varphi$ is in CNF (*Conjunctive Normal Form*); the assignment $\mu$ is initially empty, and it is updated in a stack-based manner.

In the main loop, `decide_next_branch`$(\varphi, \mu)$ chooses an unassigned literal $l$ from $\varphi$ according to some heuristic criterion, and adds it to $\mu$. (This operation is called *decision*, $l$ is called *decision literal* and the number of decision literals in $\mu$ after this operation is called the *decision level* of $l$.) In the inner loop, `deduce`$(\varphi, \mu)$ iteratively deduces literals $l$

```
1.      SatValue DPLL (formula φ, assignment μ) {
2.          while (1) {
3.              decide_next_branch(φ, μ);
4.              while (1) {
5.                  status = deduce(φ, μ);
6.                  if (status == sat)
7.                      return sat;
8.                  else if (status == conflict) {
9.                      blevel = analyze_conflict(φ, μ);
10.                     if (blevel == 0) return unsat;
11.                     else backtrack(blevel,φ, μ);
12.                 }
13.                 else break;
14.      }}}
```

Figure 1: Schema of a modern SAT solver engine based on DPLL.

deriving from the current assignment and updates $\varphi$ and $\mu$ accordingly; this step is repeated until either $\mu$ satisfies $\varphi$, or $\mu$ falsifies $\varphi$, or no more literals can be deduced, returning sat, conflict and unknown respectively. (The iterative application of Boolean deduction steps in deduce is also called *Boolean Constraint Propagation, BCP.*) In the first case, DPLL returns sat. If the second case, analyze_conflict($\varphi,\mu$) detects the subset $\eta$ of $\mu$ which caused the conflict (*conflict set*) and the decision level blevel to backtrack. If blevel == 0, then a conflict exists even without branching, so that DPLL returns unsat. Otherwise, backtrack(blevel,$\varphi,\mu$) adds the clause $\neg\eta$ to $\varphi$ (*learning*) and backtracks up to blevel (*backjumping*), updating $\varphi$ and $\mu$ accordingly. In the third case, DPLL exits the inner loop, looking for the next decision.

Notably, modern DPLL implementations implement techniques, like the two-watched-literal scheme, which allow for extremely efficient handling of BCP (Moskewicz et al., 2001; Zhang & Malik, 2002). Old versions of DPLL used to implement also the *Pure-Literal Rule (PLR)* (Davis et al., 1962): when one proposition occurs only positively (resp. negatively) in the formula, it can be safely assigned to *true* (resp. *false*). Modern DPLL implementations, however, often do not implement it anymore due to its computational cost. For a much deeper description of modern DPLL-based SAT solvers, we refer the reader to the literature (e.g., Zhang & Malik, 2002).

## 3. The Basic Encoding

We borrow some notation from the *Single Step Tableau (SST)* framework (Massacci, 2000; Donini & Massacci, 2000). We represent uniquely states in $\mathcal{M}$ as labels $\sigma$, represented as non empty sequences of integers $1.n_1^{r_1}.n_2^{r_2}. \dots .n_k^{r_k}$, s.t. the label 1 represents the root state, and $\sigma.n^r$ represents the $n$-th $\mathcal{R}_r$-successor of $\sigma$ (where $r \in \{1, \dots, m\}$). With a little abuse of notation, hereafter we may say "a state $\sigma$" meaning "a state labeled by $\sigma$". We call a *labeled formula* a pair $\langle \sigma, \psi \rangle$, such that $\sigma$ is a state label and $\psi$ is a $K_m$-formula, and we call *labeled subformulas of* a labeled formula $\langle \sigma, \psi \rangle$ all the labeled formulas $\langle \sigma, \phi \rangle$ such that $\phi$ is a subformula of $\psi$.

Let $A_{\langle\ ,\ \rangle}$ be an *injective* function which maps a labeled formula $\langle \sigma, \psi \rangle$, s.t. $\psi$ is not in the form $\neg\phi$, into a Boolean variable $A_{\langle \sigma,\ \psi \rangle}$. We conventionally assume that $A_{\langle \sigma,\ \top \rangle}$ is $\top$ and $A_{\langle \sigma,\ \bot \rangle}$ is $\bot$. Let $L_{\langle \sigma,\ \psi \rangle}$ denote $\neg A_{\langle \sigma,\ \phi \rangle}$ if $\psi$ is in the form $\neg\phi$, $A_{\langle \sigma,\ \psi \rangle}$ otherwise. Given a $K_m$-formula $\varphi$, the encoder $K_m2SAT$ builds a Boolean CNF formula as follows: [3]

$$K_m2SAT(\varphi) \quad \overset{\text{def}}{=} \quad A_{\langle 1,\ \varphi \rangle} \wedge Def(1,\ \varphi) \tag{1}$$

$$Def(\sigma,\ \top) \quad \overset{\text{def}}{=} \quad \top \tag{2}$$

$$Def(\sigma,\ \bot) \quad \overset{\text{def}}{=} \quad \top \tag{3}$$

$$Def(\sigma,\ A_i) \quad \overset{\text{def}}{=} \quad \top \tag{4}$$

$$Def(\sigma,\ \neg A_i) \quad \overset{\text{def}}{=} \quad \top \tag{5}$$

$$Def(\sigma,\ \alpha) \quad \overset{\text{def}}{=} \quad (L_{\langle \sigma,\ \alpha \rangle} \rightarrow (L_{\langle \sigma,\ \alpha_1 \rangle} \wedge L_{\langle \sigma,\ \alpha_2 \rangle})) \wedge Def(\sigma,\ \alpha_1) \wedge Def(\sigma,\ \alpha_2) \tag{6}$$

$$Def(\sigma,\ \beta) \quad \overset{\text{def}}{=} \quad (L_{\langle \sigma,\ \beta \rangle} \rightarrow (L_{\langle \sigma,\ \beta_1 \rangle} \vee L_{\langle \sigma,\ \beta_2 \rangle})) \wedge Def(\sigma,\ \beta_1) \wedge Def(\sigma,\ \beta_2) \tag{7}$$

$$Def(\sigma,\ \pi^{r,j}) \quad \overset{\text{def}}{=} \quad (L_{\langle \sigma,\ \pi^{r,j} \rangle} \rightarrow L_{\langle \sigma.j,\ \pi_0^{r,j} \rangle}) \wedge Def(\sigma.j,\ \pi_0^{r,j}) \tag{8}$$

$$Def(\sigma,\ \nu^r) \quad \overset{\text{def}}{=} \quad \bigwedge_{\substack{\text{for every} \\ \langle \sigma, \pi^{r,i} \rangle}} \left( ((L_{\langle \sigma,\ \nu^r \rangle} \wedge L_{\langle \sigma,\ \pi^{r,i} \rangle}) \rightarrow L_{\langle \sigma.i,\ \nu_0^r \rangle}) \ \wedge \ Def(\sigma.i,\ \nu_0^r) \right). \tag{9}$$

Here by "$\pi^{r,j}$" we mean that $\pi^{r,j}$ is the $j$-th distinct $\pi^r$ formula labeled by $\sigma$. Notice that (6) and (7) generalize to the case of n-ary $\wedge$ and $\vee$ in the obvious way: if $\phi$ is $\bigotimes_{i=1}^n \phi_i$ s.t. $\bigotimes \in \{\wedge, \vee\}$, then $Def(\sigma,\ \phi) \overset{\text{def}}{=} (L_{\langle \sigma,\ \phi \rangle} \rightarrow \bigotimes_{i=1}^n L_{\langle \sigma,\ \phi_i \rangle}) \wedge \bigwedge_{i=1}^n Def(\sigma,\ \phi_i)$. Although conceptually trivial, this fact has an important practical consequence: in order to encode $\bigotimes_{i=1}^n \phi_i$ one needs adding only one Boolean variable rather than up to $n-1$, see Section 4.2. Notice also that in rule (9) the literals of the type $L_{\langle \sigma,\ \pi^{r,i} \rangle}$ are strictly necessary; in fact, the SAT problem must consider and encode all the possibly occuring states, but it can be the case, e.g., that a $\pi^{r,i}$ formula occurring in a disjunction is assigned to false for a particular state label $\sigma$ (which, in SAT, corresponds to assign $L_{\langle \sigma,\ \pi^{r,i} \rangle}$ to false). In this situation all the labeled formulas regarding the state label $\sigma.i$ are useless, in particular those generated by the expansion of the $\nu$ formulas interacting with $\pi^{r,i}$. [4]

We assume that the $K_m$-formulas are represented as DAGs (Direct Acyclic Graphs), so that to avoid the expansion of the same $Def(\sigma,\ \psi)$ more than once. Then the various $Def(\sigma,\ \psi)$ are expanded in a breadth-first manner wrt. the tree of labels, that is, all the possible expansions for the same (newly introduced) $\sigma$ are completed before starting the expansions for a different state label $\sigma'$, and different state label are expanded in the order they are introduced (thus all the expansions for a given state are always handled before those of any deeper state). Moreover, following what done by Massacci (2000), we assume that, for each $\sigma$, the $Def(\sigma,\ \psi)$'s are expanded in the order: $\alpha/\beta, \pi, \nu$. Thus, each $Def(\sigma,\ \nu^r)$ is expanded after the expansion of all $Def(\sigma,\ \pi^{r,i})$'s, so that $Def(\sigma,\ \nu^r)$ will generate one clause $((L_{\langle \sigma, \ \nu^r \rangle} \land L_{\langle \sigma, \ \pi^{r,i} \rangle}) \to L_{\langle \sigma.i, \ \nu_0^r \rangle})$ and one novel definition $Def(\sigma.i, \ \nu_0^r)$ for each $Def(\sigma, \ \pi^{r,i})$ expanded. [5]

---

3. We say that the formula is in CNF because we represent clauses as implications, according to the notation described at the beginning of Section 2.

4. Indeed, (9) is a finite conjunction. In fact the number of $\pi$-subformulas is obviously finite and $K_m$ benefits of the *finite-tree-model* property (see, e.g., Pan et al., 2002; Pan & Vardi, 2003).

Intuitively, it is easy to see that $K_m 2SAT(\varphi)$ mimics the construction of an SST tableau expansion (Massacci, 2000; Donini & Massacci, 2000). We have the following fact.

**Theorem 1.** *A $K_m$-formula $\varphi$ is $K_m$-satisfiable if and only if the corresponding Boolean formula $K_m 2SAT(\varphi)$ is satisfiable.*

The complete proof of Theorem 1 can be found in Appendix A.

Notice that, due to (9), the number of variables and clauses in $K_m 2SAT(\varphi)$ may grow exponentially with $depth(\varphi)$. This is in accordance to what was stated by Halpern and Moses (1992).

**Example 3.1** (NNF). Let $\varphi_{nnf}$ be $(\Diamond A_1 \lor \Diamond (A_2 \lor A_3)) \land \Box \neg A_1 \land \Box \neg A_2 \land \Box \neg A_3$. [6] It is easy to see that $\varphi_{nnf}$ is $K_1$-unsatisfiable: the $\Diamond$-atoms impose that at least one atom $A_i$ is true in at least one successor of the root state, whilst the $\Box$-atoms impose that all atoms $A_i$ are false in all successor states of the root state. $K_m 2SAT(\varphi_{nnf})$ is: [7]

| | | |
|---|---|---|
| 1. | $A_{\langle 1, \ \varphi_{nnf} \rangle}$ | *(1)* |
| 2. | $\land \ ( \ A_{\langle 1, \ \varphi_{nnf} \rangle} \to (A_{\langle 1, \ \Diamond A_1 \lor \Diamond (A_2 \lor A_3) \rangle} \land A_{\langle 1, \ \Box \neg A_1 \rangle} \land A_{\langle 1, \ \Box \neg A_2 \rangle} \land A_{\langle 1, \ \Box \neg A_3 \rangle}) \ )$ | *(6)* |
| 3. | $\land \ ( \ A_{\langle 1, \ \Diamond A_1 \lor \Diamond (A_2 \lor A_3) \rangle} \to (A_{\langle 1, \ \Diamond A_1 \rangle} \lor A_{\langle 1, \ \Diamond (A_2 \lor A_3) \rangle}) \ )$ | *(7)* |
| 4. | $\land \ ( \ A_{\langle 1, \ \Diamond A_1 \rangle} \to A_{\langle 1.1, \ A_1 \rangle} \ )$ | *(8)* |
| 5. | $\land \ ( \ A_{\langle 1, \ \Diamond (A_2 \lor A_3) \rangle} \to A_{\langle 1.2, \ A_2 \lor A_3 \rangle} \ )$ | *(8)* |
| 6. | $\land \ ( \ (A_{\langle 1, \ \Box \neg A_1 \rangle} \land A_{\langle 1, \ \Diamond A_1 \rangle}) \to \neg A_{\langle 1.1, \ A_1 \rangle} \ )$ | *(9)* |
| 7. | $\land \ ( \ (A_{\langle 1, \ \Box \neg A_2 \rangle} \land A_{\langle 1, \ \Diamond A_1 \rangle}) \to \neg A_{\langle 1.1, \ A_2 \rangle} \ )$ | *(9)* |
| 8. | $\land \ ( \ (A_{\langle 1, \ \Box \neg A_3 \rangle} \land A_{\langle 1, \ \Diamond A_1 \rangle}) \to \neg A_{\langle 1.1, \ A_3 \rangle} \ )$ | *(9)* |
| 9. | $\land \ ( \ (A_{\langle 1, \ \Box \neg A_1 \rangle} \land A_{\langle 1, \ \Diamond (A_2 \lor A_3) \rangle}) \to \neg A_{\langle 1.2, \ A_1 \rangle} \ )$ | *(9)* |
| 10. | $\land \ ( \ (A_{\langle 1, \ \Box \neg A_2 \rangle} \land A_{\langle 1, \ \Diamond (A_2 \lor A_3) \rangle}) \to \neg A_{\langle 1.2, \ A_2 \rangle} \ )$ | *(9)* |
| 11. | $\land \ ( \ (A_{\langle 1, \ \Box \neg A_3 \rangle} \land A_{\langle 1, \ \Diamond (A_2 \lor A_3) \rangle}) \to \neg A_{\langle 1.2, \ A_3 \rangle} \ )$ | *(9)* |
| 12. | $\land \ ( \ A_{\langle 1.2, \ A_2 \lor A_3 \rangle} \to (A_{\langle 1.2, \ A_2 \rangle} \lor A_{\langle 1.2, \ A_3 \rangle}) \ )$ | *(7)* |

After a run of Boolean constraint propagation (BCP), 3. reduces to the implicate disjunction. If the first element $A_{\langle 1, \ \Diamond A_1 \rangle}$ is assigned to true, then by BCP we have a conflict on 4. and 6. If it is set to false, then the second element $A_{\langle 1, \ \Diamond (A_2 \lor A_3) \rangle}$ is assigned to true, and by BCP we have a conflict on 12. Thus $K_m 2SAT(\varphi_{nnf})$ is unsatisfiable. $\Diamond$

## 4. Optimizations

The basic encoding of Section 3 is rather naive, and can be much improved to many extents, in order to reduce the size of the output propositional formula, or to make it easier to solve by DPLL, or both. We distinguish two main kinds of optimizations:

---

5. In practice, even if the definition of $K_m 2SAT$ is recursive, the $Def$ expansions are performed grouped by states. More precisely, all the $Def(\sigma.n, \ \psi)$ expansions, for any formula $\psi$ and every defined $n$, are done together (in the $\alpha/\beta, \pi, \nu$ order above exposed) and necessarily after that all the $Def(\sigma, \ \varphi)$ expansions have been completed.

6. For $K_1$-formulas we omit the box and diamond indexes, i.e., we write $\Box, \Diamond$ for $\Box_1, \Diamond_1$.

7. In all examples we report at the very end of each line, i.e. after each clause, the number of the $K_m 2SAT$ encoding rule applied to generate that clause. We also drop the application of the rules (2), (3), (4) and (5).

**Preprocessing steps,** which are applied on the input modal formula before the encoding. Among them, we have *Pre-conversion into BNF* (Section 4.1), *Atom Normalization* (Section 4.2), *Box Lifting* (Section 4.3), and *Controlled Box Lifting* (Section 4.4).

**On-the-fly simplification steps,** which are applied to the Boolean formula under construction. Among them, we have *On-the-fly Boolean Simplification* and *Truth Propagation Through Boolean Operators* (Section 4.5) and *Truth Propagation Through Modal Operators* (Section 4.6), *On-the-fly Pure-Literal Reduction* (Section 4.7), and *On-the-fly Boolean Constraint Propagation* (Section 4.8).

We analyze these techniques in detail.

### 4.1 Pre-conversion into BNF

Many systems use to pre-convert the input $K_m$-formulas into NNF (e.g., Baader et al., 1994; Massacci, 2000). In our approach, instead, we pre-convert them into BNF (like, e.g., Giunchiglia & Sebastiani, 1996; Pan et al., 2002). For our approach, the advantage of the latter representation is that, when one $\Box_r\psi$ occurs both positively and negatively (like, e.g., in $(\Box_r\psi \lor ...) \land (\neg\Box_r\psi \lor ...)$), then both occurrences of $\Box_r\psi$ are labeled by the same Boolean atom $A_{\langle\sigma,\ \Box_r\psi\rangle}$, and hence they are always assigned the same truth value by DPLL. With NNF, instead, the negative occurrence $\neg\Box_r\psi$ is rewritten into $\Diamond_r(nnf(\neg\psi))$, so that two distinct Boolean atoms $A_{\langle\sigma,\ \Box_r(nnf(\psi))\rangle}$ and $A_{\langle\sigma,\ \Diamond_r(nnf(\neg\psi))\rangle}$ are generated; DPLL can assign them the same truth value, creating a hidden conflict which may require some extra Boolean search to reveal. [8]

**Example 4.1** (BNF). We consider the BNF variant of the $\varphi_{nnf}$ formula of Example 3.1, $\varphi_{bnf} = (\neg\Box\neg A_1 \lor \neg\Box(\neg A_2 \land \neg A_3)) \land \Box\neg A_1 \land \Box\neg A_2 \land \Box\neg A_3$. As before, it is easy to see that $\varphi_{bnf}$ is $K_1$-unsatisfiable. $K_m2SAT(\varphi_{bnf})$ is: [9]

| | | |
|---|---|---|
| 1. | $A_{\langle 1,\ \varphi_{bnf}\rangle}$ | *(1)* |
| 2. | $\land\ (\ A_{\langle 1,\ \varphi_{bnf}\rangle} \rightarrow (A_{\langle 1,\ (\neg\Box\neg A_1 \lor \neg\Box(\neg A_2 \land \neg A_3))\rangle} \land A_{\langle 1,\ \Box\neg A_1\rangle} \land A_{\langle 1,\ \Box\neg A_2\rangle} \land A_{\langle 1,\ \Box\neg A_3\rangle})\ )$ | *(6)* |
| 3. | $\land\ (\ A_{\langle 1,\ (\neg\Box\neg A_1 \lor \neg\Box(\neg A_2 \land \neg A_3))\rangle} \rightarrow (\neg A_{\langle 1,\ \Box\neg A_1\rangle} \lor \neg A_{\langle 1,\ \Box(\neg A_2 \land \neg A_3)\rangle})\ )\ )$ | *(7)* |
| 4. | $\land\ (\ \neg A_{\langle 1,\ \Box\neg A_1\rangle} \rightarrow A_{\langle 1.1,\ A_1\rangle}\ )$ | *(8)* |
| 5. | $\land\ (\ \neg A_{\langle 1,\ \Box(\neg A_2 \land \neg A_3)\rangle} \rightarrow \neg A_{\langle 1.2,\ (\neg A_2 \land \neg A_3)\rangle}\ )$ | *(8)* |
| 6. | $\land\ (\ (A_{\langle 1,\ \Box\neg A_1\rangle} \land \neg A_{\langle 1,\ \Box\neg A_1\rangle}) \rightarrow \neg A_{\langle 1.1,\ A_1\rangle}\ )$ | *(9)* |
| 7. | $\land\ (\ (A_{\langle 1,\ \Box\neg A_2\rangle} \land \neg A_{\langle 1,\ \Box\neg A_1\rangle}) \rightarrow \neg A_{\langle 1.1,\ A_2\rangle}\ )$ | *(9)* |
| 8. | $\land\ (\ (A_{\langle 1,\ \Box\neg A_3\rangle} \land \neg A_{\langle 1,\ \Box\neg A_1\rangle}) \rightarrow \neg A_{\langle 1.1,\ A_3\rangle}\ )$ | *(9)* |
| 9. | $\land\ (\ (A_{\langle 1,\ \Box\neg A_1\rangle} \land \neg A_{\langle 1,\ \Box(\neg A_2 \land \neg A_3)\rangle}) \rightarrow \neg A_{\langle 1.2,\ A_1\rangle}\ )$ | *(9)* |
| 10. | $\land\ (\ (A_{\langle 1,\ \Box\neg A_2\rangle} \land \neg A_{\langle 1,\ \Box(\neg A_2 \land \neg A_3)\rangle}) \rightarrow \neg A_{\langle 1.2,\ A_2\rangle}\ )$ | *(9)* |
| 11. | $\land\ (\ (A_{\langle 1,\ \Box\neg A_3\rangle} \land \neg A_{\langle 1,\ \Box(\neg A_2 \land \neg A_3)\rangle}) \rightarrow \neg A_{\langle 1.2,\ A_3\rangle}\ )$ | *(9)* |
| 12. | $\land\ (\ \neg A_{\langle 1.2,\ (\neg A_2 \land \neg A_3)\rangle} \rightarrow (A_{\langle 1.2,\ A_2\rangle} \lor A_{\langle 1.2,\ A_3\rangle})\ )$ | *(7)* |

Unlike with the NNF formula $\varphi_{nnf}$ in Example 3.1, $K_m2SAT(\varphi_{bnf})$ is found unsatisfiable directly by BCP. In fact, the unit-propagation of $A_{\langle 1,\ \Box\neg A_1\rangle}$ from 2. causes $\neg A_{\langle 1,\ \Box\neg A_1\rangle}$ in

---

8. Notice that this consideration holds for every representation involving both boxes and diamonds; we refer to NNF simply because it is the most popular of these representations.

9. Notice that the valid clause 6. can be dropped. See the explanation in Section 4.5.

3. to be false, so that one of the two (unsatisfiable) branches induced by the disjunction is cut a priori. With $\varphi_{nnf}$, $K_m2SAT$ does not recognize $\Box\neg A_1$ and $\Diamond A_1$ to be one the negation of the other, so that two distinct atoms $A_{\langle 1,\ \Box\neg A_1\rangle}$ and $A_{\langle 1,\ \Diamond A_1\rangle}$ are generated. Hence $A_{\langle 1,\ \Box\neg A_1\rangle}$ and $A_{\langle 1,\ \Diamond A_1\rangle}$ cannot be recognized by DPLL to be one the negation of the other, s.t. DPLL may need exploring one Boolean branch more. $\Diamond$

In the following we will assume the formulas are in BNF (although most of the optimizations which follow work also for other representations).

## 4.2 Normalization of Modal Atoms

One potential source of inefficiency for DPLL-based procedures is the occurrence in the input formula of semantically-equivalent though syntactically-different modal atoms $\psi'$ and $\psi''$ (e.g., $\Box_1(A_1 \vee A_2)$ and $\Box_1(A_2 \vee A_1)$), which are not recognized as such by $K_m2SAT$. This causes the introduction of duplicated Boolean atoms $A_{\langle \sigma,\ \psi'\rangle}$ and $A_{\langle \sigma,\ \psi''\rangle}$ and —much worse— of duplicated subformulas $Def(\sigma, \psi')$ and $Def(\sigma, \psi'')$. This fact can have very negative consequences, in particular when $\psi'$ and $\psi''$ occur with negative polarity, because this causes the creation of distinct versions of the same successor states, and the duplication of whole parts of the output formula.

**Example 4.2.** Consider the $K_m$-formula $(\phi_1 \vee \neg\Box_1(A_2 \vee A_1)) \wedge (\phi_2 \vee \neg\Box_1(A_1 \vee A_2)) \wedge \phi_3$, s.t. $\phi_1$, $\phi_2$, $\phi_3$ are possibly-big $K_m$-formulas. Then $K_m2SAT$ creates two distinct atoms $A_{\langle 1,\ \neg\Box_1(A_2 \vee A_1)\rangle}$ and $A_{\langle 1,\ \neg\Box_1(A_1 \vee A_2)\rangle}$ and two distinct formulas $Def(1, \neg\Box_1(A_2 \vee A_1))$ and $Def(1, \neg\Box_1(A_1 \vee A_2))$. The latter will cause the creation of two distinct states 1.1 and 1.2. Thus, the recursive expansion of all $\Box_1$-formulas occurring positively in $\phi_1$, $\phi_2$, $\phi_3$ will be duplicated for these two states. $\Diamond$

In order to cope with this problem, as done by Giunchiglia and Sebastiani (1996), we apply some normalization steps to modal atoms with the intent of rewriting as many as possible syntactically-different but semantically-equivalent modal atoms into syntactically-identical ones. This can be achieved by a recursive application of some simple validity-preserving rewriting rules.

**Sorting:** modal atoms are internally sorted according to some criterion, so that atoms which are identical modulo reordering are rewritten into the same atom (e.g., $\Box_i(\varphi_2 \vee \varphi_1)$ and $\Box_i(\varphi_1 \vee \varphi_2)$ are both rewritten into $\Box_i(\varphi_1 \vee \varphi_2)$).

**Flattening:** the associativity of $\wedge$ and $\vee$ is exploited and combinations of $\wedge$'s or $\vee$'s are "flattened" into n-ary $\wedge$'s or $\vee$'s respectively (e.g., $\Box_i(\varphi_1 \vee (\varphi_2 \vee \varphi_3))$ and $\Box_i((\varphi_1 \vee \varphi_2) \vee \varphi_3)$ are both rewritten into $\Box_i(\varphi_1 \vee \varphi_2 \vee \varphi_3)$).

Flattening has also the advantage of reducing the number of novel atoms introduced in the encoding, as a consequence of the fact noticed in Section 3. One possible drawback of this technique is that it can reduce the sharing of subformulas (e.g., with $\Box_i((\varphi_1 \vee \varphi_2) \vee \varphi_3)$ and $\Box_i((\varphi_1 \vee \varphi_2) \vee \varphi_4)$, the common part is no more shared). However, we have empirically experienced that this drawback is negligible wrt. the advantages of flattening.

### 4.3 Box Lifting

As second preprocessing the $K_m$-formula can also be rewritten by recursively applying the $K_m$-validity-preserving "box lifting rules":

$$(\Box_r \varphi_1 \wedge \Box_r \varphi_2) \implies \Box_r(\varphi_1 \wedge \varphi_2), \quad (\neg\Box_r \varphi_1 \vee \neg\Box_r \varphi_2) \implies \neg\Box_r(\varphi_1 \wedge \varphi_2). \quad (10)$$

This has the potential benefit of reducing the number of $\pi^r$ formulas, and hence the number of labels $\sigma.i$ to take into account in the expansion of the $Def(\sigma, \nu^r)$'s (9). We call *lifting* this preprocessing.

**Example 4.3** (Box lifting). If we apply the rules (10) to the formula of Example 4.1, then we have $\varphi_{bnflift} = \neg\Box(\neg A_1 \wedge \neg A_2 \wedge \neg A_3) \wedge \Box(\neg A_1 \wedge \neg A_2 \wedge \neg A_3)$. Consequently, $K_m 2SAT(\varphi_{bnflift})$ is:

1.              $A_{\langle 1, \, \varphi_{bnflift}\rangle}$                                                                              *(1)*

2.   $\wedge$  $(\, A_{\langle 1, \, \varphi_{bnflift}\rangle} \rightarrow (\neg A_{\langle 1, \, \Box(\neg A_1 \wedge \neg A_2 \wedge \neg A_3)\rangle} \wedge A_{\langle 1, \, \Box(\neg A_1 \wedge \neg A_2 \wedge \neg A_3)\rangle})\,)$      *(6)*

3.   $\wedge$  $(\, \neg A_{\langle 1, \, \Box(\neg A_1 \wedge \neg A_2 \wedge \neg A_3)\rangle} \rightarrow \neg A_{\langle 1.1, \, (\neg A_1 \wedge \neg A_2 \wedge \neg A_3)\rangle}\,)$                 *(8)*

4.   $\wedge$  $((\, A_{\langle 1, \, \Box(\neg A_1 \wedge \neg A_2 \wedge \neg A_3)\rangle} \wedge \neg A_{\langle 1, \, \Box(\neg A_1 \wedge \neg A_2 \wedge \neg A_3)\rangle}) \rightarrow A_{\langle 1.1, \, (\neg A_1 \wedge \neg A_2 \wedge \neg A_3)\rangle}\,)$   *(9)*

5.   $\wedge$  $(\, \neg A_{\langle 1.1, \, (\neg A_1 \wedge \neg A_2 \wedge \neg A_3)\rangle} \rightarrow (A_{\langle 1.1, \, A_1\rangle} \vee A_{\langle 1.1, \, A_2\rangle} \vee A_{\langle 1.1, \, A_3\rangle})\,)$         *(7)*

6.   $\wedge$  $(\, A_{\langle 1.1, \, (\neg A_1 \wedge \neg A_2 \wedge \neg A_3)\rangle} \rightarrow (\neg A_{\langle 1.1, \, A_1\rangle} \wedge \neg A_{\langle 1.1, \, A_2\rangle} \wedge \neg A_{\langle 1.1, \, A_3\rangle})\,).$           *(6)*

$K_m 2SAT(\varphi_{bnflift})$ is found unsatisfiable directly by BCP on clauses 1. and 2.. Only one successor state (1.1) is considered. Notice that 3., 4., 5. and 6. are redundant, because 1. and 2. alone are unsatisfiable. [10]                                           $\diamondsuit$

### 4.4 Controlled Box Lifting

One potential drawback of applying the lifting rules is that, by collapsing the formula $(\Box_r \varphi_1 \wedge \Box_r \varphi_2)$ into $\Box_r(\varphi_1 \wedge \varphi_2)$ and $(\neg\Box_r \varphi_1 \vee \neg\Box_r \varphi_2)$ into $\neg\Box_r(\varphi_1 \wedge \varphi_2)$, the possibility of sharing box subformulas in the DAG representation of the input $K_m$-formula is reduced.

In order to cope with this problem we provide an alternative policy for applying box lifting, that is, to apply the rules (10) only when neither box subformula occurring in the implicant in (10) has multiple occurrences. We call this policy *controlled box lifting*.

**Example 4.4** (Controlled Box Lifting). We apply *Controlled Box Lifting* to the formula of Example 4.1, then we have $\varphi_{bnfclift} = (\neg\Box\neg A_1 \vee \neg\Box(\neg A_2 \wedge \neg A_3)) \wedge \Box\neg A_1 \wedge \Box(\neg A_2 \wedge \neg A_3)$ since the rules (10) are applied among all the box subformulas except for $\Box\neg A_1$, which is

---

10. In our actual implementation, trivial cases like $\varphi_{bnflift}$ are found to be unsatisfiable directly during the construction of the DAG representations, so their encoding is never generated.

shared. It follows that $K_m2SAT(\varphi_{bnfclift})$ is:

1.   $A_{\langle 1,\ \varphi_{bnfclift}\rangle}$   *(1)*

2.   $\wedge$   $(\ A_{\langle 1,\ \varphi_{bnfclift}\rangle} \to (A_{\langle 1,\ (\neg\Box\neg A_1 \vee \neg\Box(\neg A_2 \wedge \neg A_3))\rangle} \wedge A_{\langle 1,\ \Box\neg A_1\rangle} \wedge A_{\langle 1,\ \Box(\neg A_2 \wedge \neg A_3)\rangle}\ )$   *(6)*

3.   $\wedge$   $(\ A_{\langle 1,\ (\neg\Box\neg A_1 \vee \neg\Box(\neg A_2 \wedge \neg A_3))\rangle} \to (\neg A_{\langle 1,\ \Box\neg A_1\rangle} \vee \neg A_{\langle 1,\ \Box(\neg A_2 \wedge \neg A_3)\rangle})\ )\ )$   *(7)*

4.   $\wedge$   $(\ \neg A_{\langle 1,\ \Box\neg A_1\rangle} \to A_{\langle 1.1,\ A_1\rangle}\ )$   *(8)*

5.   $\wedge$   $(\ \neg A_{\langle 1,\ \Box(\neg A_2 \wedge \neg A_3)\rangle} \to \neg A_{\langle 1.2,\ (\neg A_2 \wedge \neg A_3)\rangle}\ )$   *(8)*

6.   $\wedge$   $(\ (A_{\langle 1,\ \Box\neg A_1\rangle} \wedge \neg A_{\langle 1,\ \Box\neg A_1\rangle}) \to \neg A_{\langle 1.1,\ A_1\rangle}\ )$   *(9)*

7.   $\wedge$   $(\ (A_{\langle 1,\ \Box(\neg A_2 \wedge \neg A_3)\rangle} \wedge \neg A_{\langle 1,\ \Box\neg A_1\rangle}) \to A_{\langle 1.1,\ (\neg A_2 \wedge \neg A_3)\rangle}\ )$   *(9)*

8.   $\wedge$   $(\ (A_{\langle 1,\ \Box\neg A_1\rangle} \wedge \neg A_{\langle 1,\ \Box(\neg A_2 \wedge \neg A_3)\rangle}) \to \neg A_{\langle 1.2,\ A_1\rangle}\ )$   *(9)*

9.   $\wedge$   $(\ (A_{\langle 1,\ \Box(\neg A_2 \wedge \neg A_3)\rangle} \wedge \neg A_{\langle 1,\ \Box(\neg A_2 \wedge \neg A_3)\rangle}) \to A_{\langle 1.2,\ (\neg A_2 \wedge \neg A_3)\rangle}\ )$   *(9)*

10.   $\wedge$   $(\ A_{\langle 1.1,\ (\neg A_2 \wedge \neg A_3)\rangle} \to (\neg A_{\langle 1.1,\ A_2\rangle} \wedge \neg A_{\langle 1.1,\ A_3\rangle})\ )$   *(6)*

11.   $\wedge$   $(\ \neg A_{\langle 1.2,\ (\neg A_2 \wedge \neg A_3)\rangle} \to (A_{\langle 1.2,\ A_2\rangle} \vee A_{\langle 1.2,\ A_3\rangle})\ )$   *(7)*

12.   $\wedge$   $(\ A_{\langle 1.2,\ (\neg A_2 \wedge \neg A_3)\rangle} \to (\neg A_{\langle 1.2,\ A_2\rangle} \wedge \neg A_{\langle 1.2,\ A_3\rangle})\ )$   *(6)*

$K_m2SAT(\varphi_{bnfclift})$ is found unsatisfiable directly by BCP on clauses 1., 2. and 3.. Notice that the unit propagation of $A_{\langle 1,\ \Box\neg A_1\rangle}$ and $A_{\langle 1,\ \Box(\neg A_2 \wedge \neg A_3)\rangle}$ from 2. causes the implicate disjunction in 3. to be false.   $\diamond$

### 4.5 On-the-fly Boolean Simplification and Truth Propagation

A first straightforward on-the-fly optimization is that of applying recursively the standard rewriting rules for the Boolean simplification of the formula like, e.g.,

$$\langle\sigma,\varphi\rangle \wedge \langle\sigma,\varphi\rangle \implies \langle\sigma,\varphi\rangle, \qquad \langle\sigma,\varphi\rangle \vee \langle\sigma,\varphi\rangle \implies \langle\sigma,\varphi\rangle,$$
$$\langle\sigma,\varphi_1\rangle \wedge \langle\sigma,(\varphi_1 \vee \varphi_2)\rangle \implies \langle\sigma,\varphi_1\rangle, \qquad \langle\sigma,\varphi_1\rangle \vee \langle\sigma,(\varphi_1 \wedge \varphi_2)\rangle \implies \langle\sigma,\varphi_1\rangle,$$
$$\langle\sigma,\varphi\rangle \wedge \neg\langle\sigma,\varphi\rangle \implies \langle\sigma,\bot\rangle, \qquad \langle\sigma,\varphi\rangle \vee \neg\langle\sigma,\varphi\rangle \implies \langle\sigma,\top\rangle,$$
$$...,$$

and for the propagation of truth/falsehood through Boolean operators like, e.g.,

$$\neg\langle\sigma,\bot\rangle \implies \langle\sigma,\top\rangle, \qquad \neg\langle\sigma,\top\rangle \implies \langle\sigma,\bot\rangle,$$
$$\langle\sigma,\varphi\rangle \wedge \langle\sigma,\top\rangle \implies \langle\sigma,\varphi\rangle, \qquad \langle\sigma,\varphi\rangle \wedge \langle\sigma,\bot\rangle \implies \langle\sigma,\bot\rangle,$$
$$\langle\sigma,\varphi\rangle \vee \langle\sigma,\top\rangle \implies \langle\sigma,\top\rangle, \qquad \langle\sigma,\varphi\rangle \vee \langle\sigma,\bot\rangle \implies \langle\sigma,\varphi\rangle,$$
$$....$$

**Example 4.5.** If we consider the $K_m$-formula $\varphi_{bnflift} = \neg\Box(\neg A_1 \wedge \neg A_2 \wedge \neg A_3) \wedge \Box(\neg A_1 \wedge \neg A_2 \wedge \neg A_3)$ of Example 4.3 and we apply the Boolean simplification rule $\langle\sigma,\varphi\rangle \wedge \neg\langle\sigma,\varphi\rangle \implies \langle\sigma,\bot\rangle$, then $\langle\sigma,\varphi_{bnflift}\rangle$ is simplified into $\langle\sigma,\bot\rangle$.   $\diamond$

One important subcase of on-the-fly Boolean simplification avoids the useless encoding of incompatible $\pi^r$ and $\nu^r$ formulas. In BNF, in fact, the same subformula $\Box_r\psi$ may occur in the same state $\sigma$ both positively and negatively (like $\pi^r = \neg\Box_r\psi$ and $\nu^r = \Box_r\psi$). If so, $K_m2SAT$ labels both those occurrences of $\Box_r\psi$ with the same Boolean atom $A_{\langle\sigma,\ \Box_r\psi\rangle}$, and produces recursively two distinct subsets of clauses in the encoding, by applying (8) to $\neg\Box_r\psi$ and (9) to $\Box_r\psi$ respectively. However, the latter step (9) generates a *valid* clause $(A_{\langle\sigma,\ \Box_r\psi\rangle} \wedge \neg A_{\langle\sigma,\ \Box_r\psi\rangle}) \to A_{\langle\sigma.i,\ \psi\rangle}$, so that we can avoid generating it. Consequently, if $A_{\langle \sigma.i, \psi \rangle}$ no more occurs in the formula, then $Def(\sigma.i, \psi)$ should not be generated, as there is no more need of defining $\langle \sigma.i, \psi \rangle$. [11]

**Example 4.6.** If we apply this observation in the construction of the formulas of Examples 4.1 and 4.4, we have the following facts:

- In the formula $K_m 2SAT(\varphi_{bnf})$ of Example 4.1, clause 6. is valid and thus it is dropped.

- In the formula $K_m 2SAT(\varphi_{bnfclift})$ of Example 4.4, both valid clauses 6. and 9. are dropped, so that 12. is not generated. $\diamond$

Hereafter we assume that on-the-fly Boolean simplification is applied also in combination with the techniques described in the next sections.

### 4.6 On-the-fly Truth Propagation Through Modal Operators

Truth and falsehood —which can derive by the application of the techniques in Section 4.5, Section 4.7 and Section 4.8— may be propagated on-the-fly also though modal operators. First, for every $\sigma$, both positive and negative instances of $\langle \sigma, \Box_r \top \rangle$ can be safely simplified by applying the rewriting rule $\langle \sigma, \Box_r \top \rangle \Longrightarrow \langle \sigma, \top \rangle$.

Second, we notice the following fact. When we have a positive occurrence of $\langle \sigma, \neg\Box_r \bot \rangle$ for some $\sigma$ (we suppose wlog. that we have only that $\pi^r$-formula for $\sigma$), [12] by definition of (8) and (9) we have

$$Def(\sigma, \neg\Box_r\bot) = (L_{\langle \sigma, \neg\Box_r\bot \rangle} \to A_{\langle \sigma.j, \top \rangle}) \wedge Def(\sigma.j, \top), \tag{11}$$

$$Def(\sigma, \Box_r\psi) = ((L_{\langle \sigma, \Box_r\psi \rangle} \wedge L_{\langle \sigma, \neg\Box_r\bot \rangle}) \to L_{\langle \sigma.j, \psi \rangle}) \wedge Def(\sigma.j, \psi) \tag{12}$$

for some new label $\sigma.j$ and for every $\Box_r\psi$ occurring positively in $\sigma$. $Def(\sigma, \neg\Box_r\bot)$ reduces to $\top$ because both $A_{\langle \sigma.j, \top \rangle}$ and $Def(\sigma.j, \top)$ reduce to $\top$. If at least another distinct $\pi$-formula $\neg\Box_r\varphi$ occurs positively in $\sigma$, however, there is no need for the $\sigma.j$ label in (11) and (12) to be a *new* label, and we can re-use instead the label $\sigma.i$ introduced in the expansion of $Def(\sigma, \neg\Box_r\varphi)$, as follows:

$$Def(\sigma, \neg\Box_r\varphi) = (L_{\langle \sigma, \neg\Box_r\varphi \rangle} \to L_{\langle \sigma.i, \neg\varphi \rangle}) \wedge Def(\sigma.i, \neg\varphi). \tag{13}$$

Thus (11) is dropped and, for every $\langle \sigma, \Box_r\psi \rangle$ occurring positively, we write:

$$Def(\sigma, \Box_r\psi) = ((L_{\langle \sigma, \Box_r\psi \rangle} \wedge L_{\langle \sigma, \neg\Box_r\bot \rangle}) \to L_{\langle \sigma.i, \psi \rangle}) \wedge Def(\sigma.i, \psi) \tag{14}$$

instead of (12). (Notice the label $\sigma.i$ introduced in (13) rather than the label $\sigma.j$ of (11).)

This is motivated by the fact that $Def(\sigma, \neg\Box_r\bot)$ forces the existence of at least one successor of $\sigma$ but imposes no constraints on which formulas should hold there, so that we can use some other already-defined successor state, if any. This fact has the important benefit of eliminating useless successor states from the encoding.

---

11. Here the "if" is due to the fact that it may be the case that $A_{\langle \sigma.i, \psi \rangle}$ is generated anyway from the expansion of some other subformula, like, e.g., $\Box_r(\psi \vee \phi)$. If this is the case, $Def(\sigma.i, \psi)$ must be generated anyway.

12. E.g., $\neg\Box_r\bot$ may result from applying the steps of Section 4.1 and of Section 4.5 to $\neg\Box_r(\Box_r A_1 \wedge \Diamond_r \neg A_1)$.

**Example 4.7.** Let $\varphi$ be the BNF $K$-formula:

$$(\neg A_1 \vee \neg \Box A_2) \wedge (A_1 \vee \neg \Box \bot) \wedge (\neg A_1 \vee A_3) \wedge (\neg A_1 \vee \neg A_3) \wedge (A_1 \vee \Box \neg A_4) \wedge \Box A_4.$$

$\varphi$ is $K$-inconsistent, because the only possible assignment is $\{\neg A_1, \neg \Box \bot, \Box \neg A_4, \Box A_4\}$, which is $K$-inconsistent. $K_m2SAT(\varphi)$ is encoded as follows:

| | | | |
|---|---|---|---|
| 1. | | $A_{\langle 1, \ \varphi \rangle}$ | (1) |
| 2. | $\wedge$ | $(A_{\langle 1, \ \varphi \rangle} \rightarrow (A_{\langle 1, \ (\neg A_1 \vee \neg \Box A_2) \rangle} \wedge A_{\langle 1, \ (A_1 \vee \neg \Box \bot) \rangle} \wedge A_{\langle 1, \ (\neg A_1 \vee A_3) \rangle} \wedge$ | |
| | | $\qquad A_{\langle 1, \ (A_1 \vee \neg A_4) \rangle} \wedge A_{\langle 1, \ \Box A_4 \rangle}))$ | (6) |
| 3. | $\wedge$ | $(A_{\langle 1, \ (\neg A_1 \vee \neg \Box A_2) \rangle} \rightarrow (\neg A_{\langle 1, \ A_1 \rangle} \vee \neg A_{\langle 1, \ \Box A_2 \rangle}))$ | (7) |
| 4. | $\wedge$ | $(A_{\langle 1, \ (A_1 \vee \neg \Box \bot) \rangle} \rightarrow (A_{\langle 1, \ A_1 \rangle} \vee \neg A_{\langle 1, \ \Box \bot \rangle}))$ | (7) |
| 5. | $\wedge$ | $(A_{\langle 1, \ (\neg A_1 \vee A_3) \rangle} \rightarrow (\neg A_{\langle 1, \ A_1 \rangle} \vee A_{\langle 1, \ A_3 \rangle}))$ | (7) |
| 6. | $\wedge$ | $(A_{\langle 1, \ (\neg A_1 \vee \neg A_3) \rangle} \rightarrow (\neg A_{\langle 1, \ A_1 \rangle} \vee \neg A_{\langle 1, \ A_3 \rangle}))$ | (7) |
| 7. | $\wedge$ | $(A_{\langle 1, \ (A_1 \vee \Box \neg A_4) \rangle} \rightarrow (A_{\langle 1, \ A_1 \rangle} \vee A_{\langle 1, \ \Box \neg A_4 \rangle}))$ | (7) |
| 8. | $\wedge$ | $(\neg A_{\langle 1, \ \Box A_2 \rangle} \rightarrow \neg A_{\langle 1.1, \ A_2 \rangle})$ | (8) |
| 9. | $\wedge$ | $((A_{\langle 1, \ \Box \neg A_4 \rangle} \wedge \neg A_{\langle 1, \ \Box A_2 \rangle}) \rightarrow \neg A_{\langle 1.1, \ A_4 \rangle})$ | (9) |
| 10. | $\wedge$ | $((A_{\langle 1, \ \Box A_4 \rangle} \wedge \neg A_{\langle 1, \ \Box A_2 \rangle}) \rightarrow A_{\langle 1.1, \ A_4 \rangle})$ | (9) |
| 11. | $\wedge$ | $(\neg A_{\langle 1, \ \Box \bot \rangle} \rightarrow \neg A_{\langle 1.1, \ \bot \rangle})$ | (8) |
| 12. | $\wedge$ | $((A_{\langle 1, \ \Box \neg A_4 \rangle} \wedge \neg A_{\langle 1, \ \Box \bot \rangle}) \rightarrow \neg A_{\langle 1.1, \ A_4 \rangle})$ | (9) |
| 13. | $\wedge$ | $((A_{\langle 1, \ \Box A_4 \rangle} \wedge \neg A_{\langle 1, \ \Box \bot \rangle}) \rightarrow A_{\langle 1.1, \ A_4 \rangle})$ | (9) |

Clause 11. is then simplified into $\top$. (In a practical implementation it is not even generated.) Notice that in clauses 11., 12. and 13. it is used the label 1.1 of clauses 8., 9. and 10. rather than a new label 1.2. Thus, only one successor label is generated.

When DPLL is run on $K_m2SAT(\varphi)$, by BCP 1. and 2. are immediately satisfied and the implicants are removed from 3., 4., 5., 6.. Thanks to 5. and 6., $A_{\langle 1, \ A_1 \rangle}$ can be assigned only to false, which causes 3. to be satisfied and forces the assignment of the literals $\neg A_{\langle 1, \ \Box \bot \rangle}$, $A_{\langle 1, \ \Box \neg A_4 \rangle}$ by BCP on 3. and 7. and hence of $\neg A_{\langle 1.1, \ \bot \rangle}$, $\neg A_{\langle 1.1, \ A_4 \rangle}$ and $A_{\langle 1.1, \ A_4 \rangle}$ by BCP on 12. and 13., causing a contradiction. $\diamond$

It is worth noticing that (14) is strictly necessary for the correctness of the encoding even when another $\pi$-formula occurs in $\sigma$. (E.g., in Example 4.7, without 12. and 13. the formula $K_m2SAT(\varphi)$ would become satisfiable because $A_{\langle 1, \ \Box A_2 \rangle}$ could be safely be assigned to true by DPLL, which would satisfy 8., 9. and 10..)

Hereafter we assume that this technique is applied also in combination with the techniques described in Section 4.5 and in the next sections.

## 4.7 On-the-fly Pure-Literal Reduction

Another technique, evolved from that proposed by Pan and Vardi (2003), applies Pure-Literal Reduction (PLR) on-the-fly during the construction of $K_m2SAT(\varphi)$. When for a label $\sigma$ all the clauses containing atoms in the form $A_{\langle \sigma, \ \psi \rangle}$ have been generated, if some of them occurs only positively [resp. negatively], then it can be safely assigned to true [resp. to false], and hence the clauses containing $A_{\langle \sigma, \ \psi \rangle}$ can be dropped. [13] As a consequence, some other atom $A_{\langle \sigma, \, \psi' \rangle}$ can become pure, so that the process is repeated until a fixpoint is reached.

---

13. In our actual implementation this reduction is performed directly within an intermediate data structure, so that these clauses are never generated.

**Example 4.8.** Consider the formula $\varphi_{bnf}$ of Example 4.1. During the construction of $K_m 2SAT(\varphi_{bnf})$, after 1.-8. are generated, no more clause containing atoms in the form $A_{\langle 1.1, \, \psi \rangle}$ is to be generated. Then we notice that $A_{\langle 1.1, \, A_2 \rangle}$ and $A_{\langle 1.1, \, A_3 \rangle}$ occur only negatively, so that they can be safely assigned to false. Therefore, 7. and 8. can be safely dropped. Same discourse applies lately to $A_{\langle 1.2, \, A_1 \rangle}$ and 9.. The resulting formula is found inconsistent by BCP. (In fact, notice from Example 4.1 that the atoms $A_{\langle 1.1, \, A_2 \rangle}$, $A_{\langle 1.1, \, A_3 \rangle}$, and $A_{\langle 1.2, \, A_1 \rangle}$ play no role in the unsatisfiability of $K_m 2SAT(\varphi_{bnf})$.) $\qquad\qquad \diamond$

We remark the differences between PLR and the Pure-Literal Reduction technique proposed by Pan and Vardi (2003). In KBDD (Pan et al., 2002; Pan & Vardi, 2003), the Pure-Literal Reduction is a preprocessing step which is applied to the input modal formula, either at global level (i.e. looking for pure-polarity primitive propositions for the whole formula) or, more effectively, at different modal depths (i.e. looking for pure-polarity primitive propositions for the subformulas at the same nesting level of modal operators).

Our technique is much more fine-grained, as PLR is applied on-the-fly with a single-state granularity, obtaining a much stronger reduction effect.

**Example 4.9.** Consider again the BNF $K_m$-formula $\varphi_{bnf}$ discussed in Examples 4.1 and 4.8: $\varphi_{bnf} = (\neg \Box \neg A_1 \lor \neg \Box (\neg A_2 \land \neg A_3)) \land \neg \Box A_1 \land \Box \neg A_2 \land \Box \neg A_3$. It is immediate to see that all primitive propositions $A_1$, $A_2$, $A_3$ occur at every modal depth with both polarities, so that the technique of Pan and Vardi (2003) produces no effect on this formula. $\qquad \diamond$

## 4.8 On-the-fly Boolean Constraint Propagation

One major problem of the basic encoding of Section 3 is that it is "purely-syntactic", that is, it does not consider the possible truth values of the subformulas, and the effect of their propagation through the Boolean and modal connectives. In particular, $K_m 2SAT$ applies (8) [resp. (9)] to *every* $\pi$-subformula [resp. $\nu$-subformula], regardless the fact that the truth values which can be deterministically assigned to the labeled subformulas of $\langle 1, \varphi \rangle$ may allow for dropping some labeled $\pi$-/$\nu$-subformulas, and thus prevent the need of encoding them.

One solution to this problem is that of applying Boolean Constraint Propagation (BCP) on-the-fly during the construction of $K_m 2SAT(\varphi)$, starting from the fact that $A_{\langle 1, \, \varphi \rangle}$ must be true. If a contradiction is found, then $K_m 2SAT(\varphi)$ is unsatisfiable, so that the formula is not expanded any further, and the encoder returns the formula "$\bot$". [14] When BCP allows for dropping one implication in (6)-(9) without assigning some of its implicate literals, namely $L_{\langle \sigma, \, \psi_i \rangle}$, then $\langle \sigma, \psi_i \rangle$ needs not to be defined, so that $Def(\sigma, \, \psi_i)$ must not be expanded. [15] Importantly, dropping $Def(\sigma, \, \pi^{r,j})$ for some $\pi$-formula $\langle \sigma, \pi^{r,j} \rangle$ prevents generating the label $\sigma.j$ (8) and all its successor labels $\sigma.j.\sigma'$ (corresponding to the subtree of states rooted in $\sigma.j$), so that all the corresponding labeled subformulas are not encoded.

---

14. For the sake of compatibility with standard SAT solvers, our actual implementation returns the formula $A_1 \land \neg A_1$.

15. Here we make the same consideration as in Footnote 11: if $L_{\langle \sigma.j, \, \psi \rangle}$ is generated also from the expansion of some other subformula, (e.g., $\Box_r(\psi \lor \phi)$), then (another instance of) $Def(\sigma.i, \, \psi)$ must be generated anyway.

**Example 4.10.** Consider Example 4.1, and suppose we apply on-the-fly BCP. During the construction of 1., 2. and 3. in $K_m2SAT(\varphi_{bnf})$, the atoms $A_{\langle 1, \varphi_{bnf}\rangle}$, $A_{\langle 1, (\neg\Box\neg A_1\vee\neg\Box(\neg A_2\wedge\neg A_3))\rangle}$, $A_{\langle 1, \Box\neg A_1\rangle}$, $A_{\langle 1, \Box\neg A_2\rangle}$ and $A_{\langle 1, \Box\neg A_3\rangle}$ are deterministically assigned to true by BCP. This causes the removal from 3. of the first-implied disjunct $\neg A_{\langle 1, \Box\neg A_1\rangle}$, so that there is no need to generate $Def(1, \neg\Box\neg A_1)$, and hence label 1.1. is not defined and 4. is not generated. While building 5., $A_{\langle 1.2, (\neg A_2\wedge\neg A_3)\rangle}$, is unit-propagated. As label 1.1. is not defined, 6., 7. and 8. are not generated. Then during the construction of 5., 9., 10., 11. and 12., by applying BCP a contradiction is found, so that $K_m2SAT(\varphi)$ is $\bot$.

An analogous situation happens with $\varphi_{bnflift}$ in Example 4.3: while building 1. and 2. a contradiction is found by BCP, s.t. $K_m2SAT$ returns $\bot$ without expanding the formula any further. Same discourse holds for $\varphi_{bnfclift}$ in Example 4.4: while building 1., 2. and 3. a contradiction is found by BCP, s.t. $K_m2SAT$ returns $\bot$ without expanding the formula any further. $\diamond$

### 4.9 A Paradigmatic Example: Halpern & Moses Branching Formulas.

Among all optimizations described in this Section 4, on-the-fly BCP is by far the most effective. In order to better understand this fact, we consider as a paradigmatic example the branching formulas $\varphi_h^K$ by Halpern and Moses (1992, 1995) (also called "k_branch_n" in the set of benchmark formulas proposed by Heuerding and Schwendimann, 1996) and their unsatisfiable version (called "k_branch_p" in the above-mentioned benchmark suite).

Given a single modality $\Box$, an integer parameter $h$, and the primitive propositions $D_0, \ldots, D_{h+1}, P_1, \ldots, P_h$, the formulas $\varphi_h^K$ are defined as follows: [16]

$$\varphi_h^K \overset{\text{def}}{=} D_0 \wedge \neg D_1 \wedge \bigwedge_{i=0}^{h} \Box^i(depth \wedge determined \wedge branching), \tag{15}$$

$$depth \overset{\text{def}}{=} \bigwedge_{i=1}^{h+1}(D_i \rightarrow D_{i-1}), \tag{16}$$

$$determined \overset{\text{def}}{=} \bigwedge_{i=1}^{h}\left(D_i \rightarrow \left(\begin{array}{c}(P_i \rightarrow \Box(D_i \rightarrow P_i)) \wedge \\ (\neg P_i \rightarrow \Box(D_i \rightarrow \neg P_i))\end{array}\right)\right), \tag{17}$$

$$branching \overset{\text{def}}{=} \bigwedge_{i=0}^{h-1}\left((D_i \wedge \neg D_{i+1}) \rightarrow \left(\begin{array}{c}\Diamond(D_{i+1} \wedge \neg D_{i+2} \wedge P_{i+1}) \wedge \\ \Diamond(D_{i+1} \wedge \neg D_{i+2} \wedge \neg P_{i+1})\end{array}\right)\right). \tag{18}$$

A conjunction of the formulas *depth*, *determined* and *branching* is repeated at every nesting level of modal operators (i.e. at every depth): *depth* captures the relation between the $D_i$'s at every level; *determined* states that, if $P_i$ is true [false] in a state at depth $\geq i$, then it is true [false] in all the successor states of depth $\geq i$; *branching* states that, for every node at depth $i$, it is possible to find two successor states at depth $i+1$ such that $P_{i+1}$ is true in one and false in the other. For each value of the parameter $h$, $\varphi_h^K$ is K-satisfiable, and every Kripke model $M$ that satisfies it has at least $2^{h+1}-1$ states. In fact, $\varphi_h^K$ is build in such a way to force the construction of a binary-tree Kripke model of depth $h+1$, each of whose leaves encodes a distinct truth assignment to the primitive propositions $P_1, \ldots, P_h$, whilst each $D_i$ is true in all and only the states occurring at a depth $\geq i$ in the tree (and thus denotes the level of nesting).

---

16. For the sake of better readability, here we adopt the description given by Halpern and Moses (1992) without converting the formulas into BNF. This fact does not affect the discussion.

The unsatisfiable counterpart formulas proposed by Heuerding and Schwendimann (1996) (whose negations are the valid formulas called `k_branch_p` in the previously-mentioned benchmark suite, which are exposed in more details in Section 5.1.1) are obtained by conjoining to (15) the formula:

$$\Box^h P_{\lfloor \frac{h}{3} \rfloor + 1} \tag{19}$$

(where $\lfloor x \rfloor$ is the integer part of $x$) which forces the atom $P_{\lfloor \frac{h}{3} \rfloor + 1}$ to be true in all depth-$h$ states of the candidate Kripke model, which is incompatible with the fact that the remaining specifications say that it has to be false in half depth-$h$ states. [17]

These formulas are very pathological for many approaches (Giunchiglia & Sebastiani, 2000; Giunchiglia, Giunchiglia, Sebastiani, & Tacchella, 2000; Horrocks et al., 2000). In particular, before introducing on-the-fly BCP, they used to be the pet hate of our $K_m2SAT$ approach, as they caused the generation of huge Boolean formulas. In fact, due to *branching* (18), $\varphi_h^K$ contains $2h$ $\Diamond$-formulas (i.e., $\pi$-formulas) at every depth. Therefore, the $K_m2SAT$ encoder of Section 3 has to consider $1 + 2h + (2h)^2 + \ldots + (2h)^{h+1} = ((2h)^{h+2} - 1)/(2h - 1)$ distinct labels, which is about $h^{h+1}$ times the number of those labeling the states which are actually needed. (None of the optimizations of Sections 4.1-4.7 is of any help with these formulas, because neither BNF encoding nor atom normalization causes any sharing of subformulas, the formulas are already in lifted form, and no literal occurs pure. [18])

This pathological behavior can be mostly overcome by applying on-the-fly-BCP, because some truth values can be deterministically assigned to some subformulas of $\varphi_h^K$ by on-the-fly-BCP, which prevent encoding some or even most $\Box/\Diamond$-subformulas.

In fact, consider the *branching* and *determined* formulas occurring in $\varphi_h^K$ at a generic depth $d \in \{0 \ldots h\}$, which determine the states at level $d$ in the tree. As in these states $D_0, \ldots, D_d$ are forced to be true and $D_{d+1}, \ldots, D_{h+1}$ are forced to be false, then all but the $d$-th conjunct in *branching* (all conjuncts if $d = h$) are forced to be true and thus they could be dropped. Therefore, only 2 $\Diamond$-formulas per non-leaf level could be considered instead, causing the generation of $2^{h+1} - 1$ labels overall. Similarly, in all states at level $d$ the last $h - d$ conjuncts in *determined* are forced to be true and could be dropped, reducing significantly the number of $\Box$-formulas to be considered.

It is easy to see that this is exactly what happens by applying on-the-fly-BCP. In fact, suppose that the construction of $K_m2SAT(\varphi_h^K)$ has reached depth $d$ (that is, the point where for every state $\sigma$ at level $d$, the $Def(\sigma, \alpha)$'s and $Def(\sigma, \beta)$'s are expanded but no $Def(\sigma, \pi)$ and $Def(\sigma, \nu)$ is expanded yet). Then, BCP deterministically assigns true to the literals $L_{\langle \sigma, D_0 \rangle}, \ldots, L_{\langle \sigma, D_d \rangle}$ and false to $L_{\langle \sigma, D_{d+1} \rangle}, \ldots, L_{\langle \sigma, D_{h+1} \rangle}$, which removes all but one conjuncts in *branching*, so that only two $Def(\sigma, \pi)$'s out of $2h$ ones are actually expanded; similarly, the last $h - d$ conjuncts in *determined* are removed, so that the corresponding $Def(\sigma, \nu)$'s are not expanded.

---

17. Heuerding and Schwendimann do not explain the choice of the index "$\lfloor \frac{h}{3} \rfloor + 1$". We understand that also other choices would have done the job.

18. More precisely, only one literal, $\neg D_{h+1}$, occurs pure in *branching*, but assigning it plays no role in simplifying the formula.

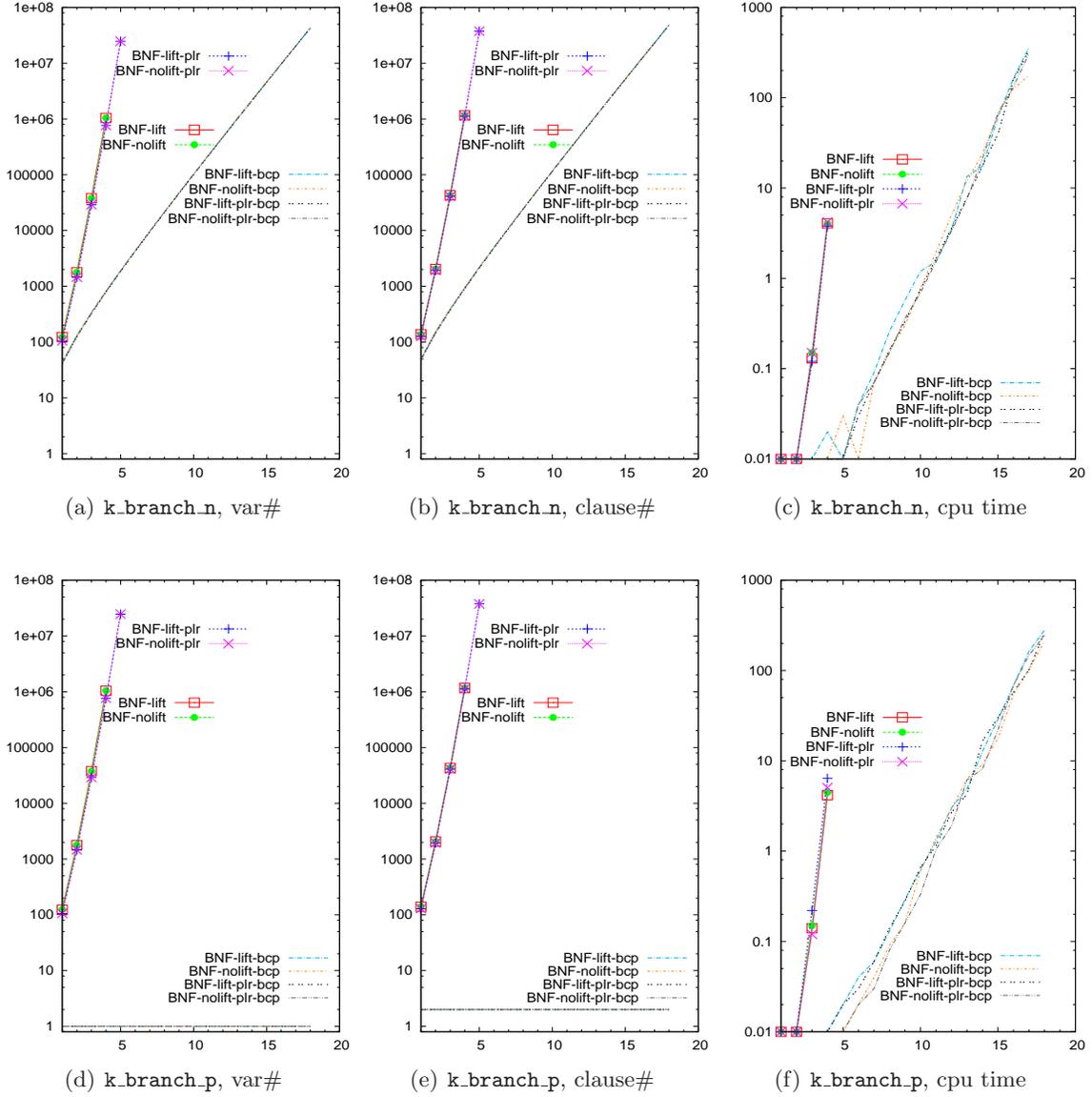

Figure 2: Empirical analysis of $K_m2SAT$ on Halpern & Moses formulas wrt. the depth parameter $h$, for different options of the encoder. 1st row: `k_branch_n`, corresponding to $K_m2SAT(\varphi_h^K)$, formulas (satisfiable); 2nd row: `k_branch_p`, corresponding to $K_m2SAT(\varphi_h^K \wedge \square^h P_{\lfloor \frac{h}{3} \rfloor+1})$, formulas (unsatisfiable). Left: number of Boolean variables; center: number of clauses; right: total CPU time requested to encoding+solving (where the solving step has been performed through Rsat). See Section 5 for more technical details.

As far as the unsatisfiable version $K_m 2SAT(\varphi_h^K \wedge \square^h P_{\lfloor \frac{h}{3} \rfloor + 1})$ is concerned, when the expansion reaches depth $h$, thanks to (19), $L_{\langle \sigma, \, P_{\lfloor \frac{h}{3} \rfloor + 1} \rangle}$ is generated and deterministically assigned to true by BCP for every depth-$h$ label $\sigma$; thanks to *determined* and *branching*, BCP assigns all literals $L_{\langle \sigma, \, P_1 \rangle}, ..., L_{\langle \sigma, \, P_h \rangle}$ deterministically, so that $L_{\langle \sigma, \, P_{\lfloor \frac{h}{3} \rfloor + 1} \rangle}$ is assigned to false for 50% of the depth-$h$ labels $\sigma$. This causes a contradiction, so that the encoder stops the expansion and returns $\bot$.

Figure 2 shows the growth in size and the CPU time required to encode and solve $K_m 2SAT(\varphi_h^K)$ (1st row) and $K_m 2SAT(\varphi_h^K \wedge \square^h P_{\lfloor \frac{h}{3} \rfloor + 1})$ (2nd row) wrt. the parameter $h$, for eight combinations of the following options of the encoder: with and without box-lifting, with and without on-the-fly PLR, with and without on-the-fly BCP. (Notice the log scale of the y axis.) In Figure 2(d) the plots of the four versions "`-xxx-bcp`" (with on-the-fly BCP) coincide with the line of value 1 (i.e, one variable) and in Figure 2(e) they coincide with an horizontal line of value 2 (i.e, two clauses), corresponding to the fact that the 1-variable/2-clause formula $A_1 \wedge \neg A_1$ is returned (see Footnote 14).

We notice a few facts. First, for both formulas, the eight plots always collapse into two groups of overlapping plots, representing the four variants with and without on-the-fly BCP respectively. This shows that box-lifting and on-the-fly PLR are almost irrelevant in the encoding of these formulas, causing just little variations in the time required by the encoder (Figures 2(c) and 2(f)); notice that enabling on-the-fly PLR alone permits to encode (but not to solve) only one problem more wrt. the versions without both on-the-fly PLR and BCP. Second, the four versions with on-the-fly-BCP always outperform of several orders magnitude these without this option, in terms of both size of encoded formulas and of CPU time required to encode and solve them. In particular, in the case of the unsatisfiable variant (Figure 2, second row) the encoder returns the $\bot$ formula, so that no actual work is required to the SAT solver (the plot of Figure 2(f) refers only to encoding time).

## 5. Empirical Evaluation

In order to verify empirically the effectiveness of this approach, we have performed a very-extensive empirical test session on about 14,000 $K_m/\mathcal{ALC}$ formulas. We have implemented the encoder $K_m 2SAT$ in C++, with some flags corresponding to the optimizations exposed in the previous section: (i) `NNF/BNF`, performing a pre-conversion into NNF/BNF before the encoding; (ii) `lift/ctrl.lift/nolift`, performing respectively Box Lifting, Controlled Box Lifting or no Box Lifting before the encoding; (iii) `plr` if on-the-fly Pure Literal Reduction is performed and (iv) `bcp` if on-the-fly Boolean Constraint Propagation is performed. The techniques introduced in Section 4.2, Section 4.5 and Section 4.6 are hardwired in the encoder. Moreover, as pre-conversion into `BNF` almost always produces smaller formulas than `NNF`, we have set the `BNF` flag as a default.

In combination with $K_m 2SAT$ we have tried several SAT solvers on our encoded formulas (including Zchaff 2004.11.15, Siege v4, BerkMin 5.6.1, MiniSat v1.13, SAT-Elite v1.0, SAT-Elite GTI 2005 `submission` [19], MiniSat 2.0 061208 and Rsat 1.03).

---

19. In the preliminary evaluation of the available SAT solvers we have also tried SAT-Elite as a preprocessor to reduce the size of the SAT formula generated by $K_m 2SAT$ without the `bcp` option before to solve it. However, even if the preprocessing can signinificantly reduce the size of the formula, it has turned out

After a preliminary evaluation and further intensive experiments we have selected Rsat `1.03` (Pipatsrisawat & Darwiche, 2006), because it produced the best overall performances on our benchmark suites (although the performance gaps wrt. other SAT tools, e.g. MiniSat 2.0, were not dramatic).

We have downloaded the available versions of the state-of-the-art tools for $K_m$-satisfiability. After an empirical evaluation [20] we have selected Racer `1-7-24` (Haarslev & Moeller, 2001) and *SAT `1.3` (Tacchella, 1999) as the best representatives of the tableaux/DPLL-based tools, Mspass `v 1.0.0t.1.3` (Hustadt & Schmidt, 1999; Hustadt et al., 1999) [21] as the best representative of the FOL-encoding approach, KBDD (unique version) (Pan et al., 2002; Pan & Vardi, 2003) [22] as the representative of the automata-theoretic approach. No representative of the CSP-based and of the inverse method approaches could be used. [23] Notice that all these tools but Racer are experimental tools, as far as $K_m2SAT$ which is a prototype, and many of them (e.g. *SAT and KBDD) are no longer maintained.

Finally, as representative of the QBF-encoding approach, we have selected the K-QBF translator (Pan & Vardi, 2003) combined with the sKizzo version `0.8.2` QBF solver (Benedetti, 2005), which turned out to be by far [24] the best QBF solver on our benchmarks among the freely-available QBF solvers from the QBF2006 competition (Narizzano, Pulina, & Tacchella, 2006). (In our evaluation we have considered the tools : 2clsQ, SQBF, preQuantor—i.e. preQuel +Quantor— Quantor `2.11`, and Semprop `010604`.)

All tests presented in this section have been performed on a two-processor Intel Xeon 3.0GHz computer, with 1 MByte Cache each processor, 4 GByte RAM, with Red Hat Linux 3.0 Enterprise Server, where four processes can run in parallel. When reporting the results for one $K_m2SAT$ +Rsat version, the CPU times reported are the sums of both

---

that this preprocessing is too time-expensive and that the overall time spent for preprocessing and then solving the reduced problem is higher than that solving directly the original encoded SAT formula.

20. As we did for the selection of the SAT solver, in order to select the tools to be used in the empirical evaluation, we have performed a preliminary evaluation on the smaller benchmark suites (i.e. the LWB and, sometimes, the TANCS 2000 ones; see later). Importantly, from this preliminary evaluation Racer turned out to be definitely more efficient than FaCT++, being able to solve more problems in less time. Also, in order to meet the reviewers' suggestions, we repeated this preliminary evaluation with the latest versions of FaCT++ (`v1.2.3`, March 5th, 2009) and the same version of Racer used in this paper. In this evaluation Racer solves ten more problems than FaCT++ on the LWB benchmark, and over than one hundred of problems more than FaCT++ on the whole TANCS 2000 suite. Also on $\Box_m$-CNF random problems Racer outperforms FaCT++. (We include in the online appendix the plots of this comparison between Racer and FaCT++.)

21. We have run Mspass with the options `-EMLTranslation=2 -EMLFuncNary=1 -Sorts=0 -CNFOptSkolem=0 -CNFStrSkolem=0 -Select=2 -Split=-1 -DocProof=0 -PProblem=0 -PKept=0 -PGiven=0`, which are suggested for $K_m$-formulas in the Mspass README file. We have also tried other options, but the former gave the best performances.

22. KBDD has been recompiled to be run with an increased internal memory bound of 1 GB.

23. At the moment KӾ is not freely available, and we failed in the attempt of obtaining it from the authors. KCSP is a prolog piece of software, which is difficult to compare in performances wrt. other optimized tools on a common platform; moreover, KCSP is no more maintained since 2005, and it is not competitive wrt. state-of-the-art tools (Brand, 2008). Other tools like leanK, $\Box$KE, LWB, Kris are not competitive with the ones listed above (Horrocks et al., 2000). KSAT (Giunchiglia & Sebastiani, 1996, 2000; Giunchiglia et al., 2000) has been reimplemented into *SAT.

24. Unlike with the choice of SAT solver, the performance gaps from the best choice and the others were very significant: e.g., in the LWB benchmark (see later), sKizzo was able to solve nearly 90 problems more than its best QBF competitor.

the encoding and RSAT solving times. When reporting the results for K-QBF +sKIZZO, the CPU times reported are only due to sKIZZO because the time spent by the K-QBF converter is negligible.

We anticipate that, for all formulas of all benchmark suites, all tools under test —i.e. all the variants of $K_m2SAT$ +RSAT and all the state-of-the-art $K_m$-satisfiability solvers— agreed on the satisfiability/unsatisfiability result when terminating within the timeout.

*Remark* 1. Due to the big number of empirical tests performed and to the huge amount of data plotted, and due to limitations in size, and in order to to make the plots clearly distinguishable in the figures, we have limited the number of plots included in the following part of the paper, considering only the most meaningful ones and those regarding the most challenging benchmark problems faced. For the sake of the reader's convenience, however, full-size versions of all plots and many other plots regarding the not-exposed results (also on the easier problems), are available in the online appendix, together with the files with all data. When discussing the empirical evaluation we may include in our considerations also these results.

## 5.1 Test Description

We have performed our empirical evaluation on three different well-known benchmarks suites of $K_m/\mathcal{ALC}$ problems: the LWB (Heuerding & Schwendimann, 1996), the random $\Box_m$-CNF (Horrocks et al., 2000; Patel-Schneider & Sebastiani, 2003) and the TANCS 2000 (Massacci & Donini, 2000) benchmark suites. We are not aware of any other publicly-available benchmark suite on $K_m/\mathcal{ALC}$-satisfiability from the literature. These three groups of benchmark formulas allow us to test the effectiveness of our approach on a large number of problems of various sizes, depths, hardness and characteristics, for a total amount of about 14,000 formulas.

In particular, these benchmark formulas allow us to fairly evaluate the different tools both on the *modal* component and on the *Boolean* component of reasoning which are intrinsic in the $K_m$-satisfiability problem, as we discuss later in Section 5.4.

In the following we describe these three benchmark suites.

### 5.1.1 THE LWB BENCHMARK SUITE

As a first group of benchmark formulas we used the LWB benchmark suite used in a comparison at Tableaux'98 (Heuerding & Schwendimann, 1996). It consists of 9 classes of parametrized formulas (each in two versions, provable "_p" or not-provable "_n" [25]), for a total amount of 378 formulas. The parameter allows for creating formulas of increasing size and difficulty.

The benchmark methodology is to test formulas from each class, in increasing difficulty, until one formula cannot be solved within a given timeout, 1000 seconds in our tests. [26] The result from this class is the parameter's value of the largest (and hardest) formula that can be solved within the time limit. The parameter ranges only from 1 to 21 so that, if a system can solve all 21 instances of a class, the result is given as 21. For a discussion on this benchmark suite, we refer the reader to the work of Heuerding and Schwendimann (1996) and of Horrocks et al. (2000).

---

25. Since all tools check $K_m$-(un)satisfiability, all formulas are negated, so that the negations of the provable formulas are checked to be unsatisfiable, whilst the negation of the other formulas are checked to be satisfiable.
26. We also set a 1 GB file-size limit for the encoding produced by $K_m2SAT$.

### 5.1.2 THE RANDOM $\Box_m$-CNF BENCHMARK SUITE

As a second group of benchmark formulas, we have selected the random $\Box_m$-CNF testbed described by Horrocks et al. (2000), and Patel-Schneider and Sebastiani (2003). This is a generalization of the well-known random k-SAT test methods, and is the final result of a long discussion in the communities of modal and description logics on how to to obtain significant and flawless random benchmarks for modal/description logics (Giunchiglia & Sebastiani, 1996; Hustadt & Schmidt, 1999; Giunchiglia et al., 2000; Horrocks et al., 2000; Patel-Schneider & Sebastiani, 2003).

In the $\Box_m$-CNF test methodology, a $\Box_m$-CNF formula is randomly generated according to the following parameters:

- the (maximum) modal depth $d$;

- the number of top-level clauses $L$;

- the number of literal per clause clauses $k$;

- the number of distinct propositional variables $N$;

- the number of distinct box symbols $m$;

- the percentage $p$ of purely-propositional literals in clauses occurring at depth $< d$, s.t. each clause of length $k$ contains on average $p \cdot k$ randomly-picked Boolean literals and $k - p \cdot k$ randomly-generated modal literals $\Box_r \psi$, $\neg \Box_r \psi$. [27]

(We refer the reader to the works of Horrocks et al., 2000, and Patel-Schneider & Sebastiani, 2003 for a more detailed description.)

A typical problem set is characterized by fixed values of $d$, $k$, $N$, $m$, and $p$: $L$ is varied in such a way as to empirically cover the "100% satisfiable / 100% unsatisfiable" transition. In other words, many problems with the same values of $d, k, N, m$, and $p$ but an increasing number of clauses $L$ are generated, starting from really small, typically satisfiable problems (i.e. with a probability of generating a satisfiable problem near to one) to huge problems, where the increasing interactions among the numerous clauses typically leads to unsatisfiable problems (i.e. it makes the probability of generating satisfiable problems converging to zero). Then, for each tuple of the five values in a problem set, a certain number of $\Box_m$-CNF formulas are randomly generated, and the resulting formulas are given in the input to the procedure under test, with a maximum time bound. The fraction of formulas which were solved within a given timeout, and the median/percentile values of CPU times are plotted against the ratio $L/N$. Also, the fraction of satisfiable/unsatisfiable formulas is plotted for a better understanding.

---

27. More precisely, the number of Boolean literals in a clause is $\lfloor p \cdot k \rfloor$ (resp. $\lceil p \cdot k \rceil$) with probability $\lceil p \cdot k \rceil - p \cdot k$ (resp. $1 - (\lceil p \cdot k \rceil - p \cdot k)$). Notice that typically the smaller is $p$, the harder is the problem (Horrocks et al., 2000; Patel-Schneider & Sebastiani, 2003).

Following the methodology proposed by Horrocks et al. (2000), and by Patel-Schneider and Sebastiani (2003), we have fixed $m = 1$, $k = 3$ and 100 samples per point in all tests, and we have selected two groups: an "easier" one, with $d = 1$, $p = 0.5$, $N = 6, 7, 8, 9$, $L/N = 10..60$, and a "harder" one, with $d = 2$, $p = 0.6, 0.5$, $N = 3, 4$, $L/N = 30..150$ with $p = 0.6$ and $L/N = 50..140$ with $p = 0.5$, varying the $L/N$ ratio in steps of 5, for a total amount of 13,200 formulas.

In each test, we imposed a timeout of 500 seconds per sample [28] and we calculated the number of samples which were solved within the timeout, and the 50%th and 90%th percentiles of CPU time. [29] In order to correlate the performances with the (un)satisfiability of the sample formulas, in the background of each plot we also plot the satisfiable/unsatisfiable ratio.

### 5.1.3 The TANCS 2000 Benchmark Suite

Finally, as a third group of benchmark formulas, we used the MODAL PSPACE division benchmark suite used in the comparison at TANCS 2000 (Massacci & Donini, 2000). It contains both satisfiable and unsatisfiable formulas, with scalable hardness. In this benchmark suite, which we call TANCS 2000, the formulas are constructed by translating QBF formulas into $K$ using three translation schemas, namely the Schmidt-Schauss-Smolka translation (240 problems with many different depths, from 19 to 112), the Ladner translation (240 problems, again with depths in the same range $19 - 112$), and the Halpern translation (56 problems of depth among: 20, 28, 40, 56, 80 or 112) (Massacci & Donini, 2000). As done by Massacci and Donini, we call these classes *easy*, *medium* and *hard* respectively.

All formulas from each class are tested within a timeout of 1000 seconds. [30] For each class, we report the number of solved formulas ($X$ axis) and the total (cumulative) CPU time spent for solving these formulas ($Y$ axes). For each class the results are plotted sorting the solved problems from the easiest one to the hardest one.

## 5.2 An Empirical Comparison of the Different Variants of $K_m 2SAT$

We have first evaluated the various variants of the encoding in combination with Rsat. In order to avoid considering too many combinations of the flags, we have considered the `BNF` format, and we have grouped `plr` and `bcp` into one parameter `plr-bcp`, restricting thus our investigation to 6 combinations: `BNF`, `lift`/`ctrl.lift`/`nolift`, and `plr-bcp` on/off. (We recall that the techniques introduced in Section 4.2, Section 4.5 and Section 4.6 are hardwired in the encoder.) Here we expose and analyze the results wrt. the three different suites of benchmark problems.

---

28. With also a 512 MB file-size limit for the encoding produced by $K_m 2SAT$.

29. Due to the lack of space and for the sake of clarity we won't include in the paper the 90%th percentiles plots. Further, for the same reasons, we'll skip to report the plots regarding some of the easiest class of the benchmark suite (e.g. those with $d = 1$ and lower values of $N$). All of these plots, however, can be found in the online appendix.

30. We also set a 1 GB file-size limit for the encoding produced by $K_m 2SAT$.

### 5.2.1 Results on the LWB Benchmark Suite

The results on the LWB benchmark suite are summarized in Table 1 and Figure 3.

Table 1(a) reports in the left block the indexes of the hardest formulas encoded within the file-size limit and, in the right block, those of the hardest formulas solved within the timeout by Rsat; Table 1(b) reports the numbers of variables and clauses of $K_m2SAT(\varphi)$, referring to the hardest formulas solved within the timeout by Rsat (i.e., those reported in the right block of Table 1(a)). For instance, the `BNF-ctrl.lift-plr-bcp` encoding of `k_dum_n(21)` contains $11 \cdot 10^6$ variables and $14 \cdot 10^6$ clauses; it is the hardest `k_dum_n` problem solved by Rsat with `BNF-ctrl.lift-plr-bcp` and it is the first which is not solved with `BNF-ctrl.lift`.

Looking at the numbers of cases solved in Table 1(a), we notice that the introduction of the on-the-fly Pure Literal Reduction and Boolean Constraint Propagation optimizations is really effective and produces a consistent performance enhancement (the effect of these optimizations is eye-catching in the branching formulas `k_branch_*` – see Section 4.9 – and in the `k_path_*` formulas). We also notice that `lift` sometimes introduces some slight further improvement.

The view of Tables 1(a) and 1(b) hides the actual CPU times required to encode and solve the problems. Small gaps in the numbers of Table 1(a) may correspond to big gaps in CPU time. In order to analyze also this aspect, in Figure 3 we plotted the total cumulative amount of CPU time spent by all the variants of $K_m2SAT$ +Rsat to solve all the problems of the LWB benchmark, sorted by hardness. For this plot, we also considered three more options —BNF, `lift/ctrl.lift/nolift`, with `plr` on and `bcp` off— so that to evaluate also the effect of `plr` and `bcp` separately. We notice that the plots are clearly clustered into three groups of increasing performance: `BNF-*`, `BNF-*-plr`, and `BNF-*-plr-bcp`., "`*`" representing the three options `lift/ctrl.lift/nolift`. This highlights the fact that on this suite on-the-fly Pure Literal Reduction significantly improves the performances, that on-the-fly Boolean Constraint Propagation introduces drastic improvements, and that the variations due to Box Lifting are minor wrt. the other two optimizations.

Overall, the configuration `BNF-lift-plr-bcp` turns out to be the best performer on this suite, with a tiny advantage wrt. `BNF-ctrl.lift-plr-bcp`.

### 5.2.2 Results on the Random $\Box_m$-CNF Benchmark Suite

The results on the random $\Box_m$-CNF benchmark suite are reported in Figures 4 and 5.

In Figure 4 we report the 50%-percentile CPU times required to encode and solve the formulas by the different $K_m2SAT$ +Rsat variants for the hardest benchmarks problems. We don't report the percentage of solved problems since it is always 100%, i.e. $K_m2SAT$ +Rsat terminates within the timeout for every problem in the benchmark suite.

The tests with depth $d = 1$ (see the results on the hardest problems of the class in the first row of Figure 4) are simply too easy for $K_m2SAT$ +Rsat (but not for its competitors, see Section 5.3) which solved every sample formula in less than 1 second. Although the tests exposed in the second and third row of Figure 4 are more challenging, they are all solved within the timeout as well. We have noticed also that the results are rather regular, since there are no big gaps between 50%- and 90%-percentile values.

| lifting | $K_m2SAT$, encoded | | | | | | $K_m2SAT$ + Rsat, solved | | | | | |
|---|---|---|---|---|---|---|---|---|---|---|---|---|
| | | | | plr-bcp | | | | | | plr-bcp | | |
| | no | yes | ctrl | no | yes | ctrl | no | yes | ctrl | no | yes | ctrl |
| k_branch_n | 4 | 4 | 4 | **18** | **18** | **18** | 4 | 4 | 4 | **17** | **17** | **17** |
| k_branch_p | 4 | 4 | 4 | **18** | **18** | **18** | 4 | 4 | 4 | **18** | **18** | **18** |
| k_d4_n | 8 | 8 | 8 | 8 | **9** | 8 | **8** | **8** | **8** | **8** | **8** | **8** |
| k_d4_p | **14** | **14** | **14** | 14 | 14 | 14 | **14** | **14** | **14** | 14 | 14 | 14 |
| k_dum_n | 20 | 20 | 20 | **21** | **21** | **21** | 20 | 20 | 20 | **21** | **21** | **21** |
| k_dum_p | 19 | 19 | 19 | **21** | **21** | **21** | 18 | 18 | 18 | **21** | **21** | **21** |
| k_grz_n | **21** | **21** | **21** | **21** | **21** | **21** | **21** | **21** | **21** | **21** | **21** | **21** |
| k_grz_p | **21** | **21** | **21** | **21** | **21** | **21** | **21** | **21** | **21** | **21** | **21** | **21** |
| k_lin_n | **21** | **21** | **21** | **21** | **21** | **21** | **21** | **21** | **21** | **21** | **21** | **21** |
| k_lin_p | **21** | **21** | **21** | **21** | **21** | **21** | **21** | **21** | **21** | **21** | **21** | **21** |
| k_path_n | 7 | 7 | 7 | 14 | **15** | 14 | 7 | 7 | 7 | 13 | **14** | 13 |
| k_path_p | 8 | 8 | 8 | 15 | **16** | 15 | 8 | 8 | 8 | 15 | **16** | 15 |
| k_ph_n | **21** | **21** | **21** | **21** | **21** | **21** | **21** | **21** | **21** | **21** | **21** | **21** |
| k_ph_p | **21** | **21** | **21** | **21** | **21** | **21** | 10 | **11** | 10 | 10 | 10 | **11** |
| k_poly_n | **21** | **21** | **21** | **21** | **21** | **21** | **21** | **21** | **21** | **21** | **21** | **21** |
| k_poly_p | **21** | **21** | **21** | **21** | **21** | **21** | **21** | **21** | **21** | **21** | **21** | **21** |
| k_t4p_n | **6** | **6** | **6** | **6** | **6** | **6** | 5 | **6** | 5 | **6** | **6** | **6** |
| k_t4p_p | **11** | **11** | **11** | **11** | **11** | **11** | 10 | 10 | 10 | **11** | **11** | **11** |

(a) Indexes of the hardest problems encoded (left)
and of the hardest problems solved (right).

| lifting | number of variables ($10^3$) | | | | | | number of clauses ($10^3$) | | | | | |
|---|---|---|---|---|---|---|---|---|---|---|---|---|
| | | | | plr-bcp | | | | | | plr-bcp | | |
| | no | yes | ctrl | no | yes | ctrl | no | yes | ctrl | no | yes | ctrl |
| k_branch_n | 1000 | 1000 | 1000 | 20000 | 20000 | 20000 | 1000 | 1000 | 1000 | 23000 | 23000 | 23000 |
| k_branch_p | 1000 | 1000 | 1000 | 0 | 0 | 0 | 1000 | 1000 | 1000 | 0 | 0 | 0 |
| k_d4_n | 12000 | 6000 | 12000 | 10000 | 26000 | 10000 | 17000 | 9000 | 17000 | 16000 | 43000 | 16000 |
| k_d4_p | 19000 | 18000 | 19000 | 0 | 0 | 0 | 28000 | 25000 | 28000 | 0 | 0 | 0 |
| k_dum_n | 19000 | 19000 | 19000 | 11000 | 11000 | 11000 | 23000 | 23000 | 23000 | 14000 | 14000 | 14000 |
| k_dum_p | 11000 | 11000 | 11000 | 20000 | 19000 | 20000 | 14000 | 13000 | 14000 | 26000 | 25000 | 26000 |
| k_grz_n | 10 | 10 | 10 | 5 | 5 | 5 | 10 | 10 | 10 | 6 | 6 | 6 |
| k_grz_p | 8 | 8 | 8 | 0.2 | 0.1 | 0.2 | 8 | 8 | 8 | 0.3 | 0.2 | 0.2 |
| k_lin_n | 30 | 30 | 20 | 20 | 10 | 20 | 50 | 50 | 20 | 30 | 30 | 30 |
| k_lin_p | 0 | 0 | 0 | 0 | 0 | 0 | 0 | 0 | 0 | 0 | 0 | 0 |
| k_path_n | 11000 | 12000 | 11000 | 10000 | 7000 | 10000 | 13000 | 14000 | 13000 | 14000 | 9000 | 13000 |
| k_path_p | 11000 | 12000 | 11000 | 26000 | 16000 | 26000 | 13000 | 14000 | 13000 | 35000 | 20000 | 35000 |
| k_ph_n | 50 | 300 | 50 | 50 | 300 | 50 | 50 | 300 | 50 | 50 | 600 | 50 |
| k_ph_p | 3 | 13 | 3 | 3 | 8 | 4 | 3 | 14 | 3 | 3 | 14 | 5 |
| k_poly_n | 200 | 20 | 20 | 200 | 20 | 20 | 200 | 20 | 20 | 200 | 20 | 20 |
| k_poly_p | 200 | 20 | 20 | 200 | 20 | 20 | 200 | 20 | 20 | 200 | 20 | 20 |
| k_t4p_n | 4000 | 21000 | 4000 | 17000 | 14000 | 17000 | 4000 | 22000 | 4000 | 20000 | 17000 | 20000 |
| k_t4p_p | 12000 | 10000 | 12000 | 20000 | 18000 | 20000 | 12000 | 11000 | 12000 | 24000 | 21000 | 24000 |

(b) # of variables and # of clauses of the hardest problems solved.
Note: Here "0" means that the formula is simplified into ⊥ by $K_m2SAT$.

Table 1: Comparison of the variants of $K_m2SAT$ +Rsat on the LWB benchmarks.

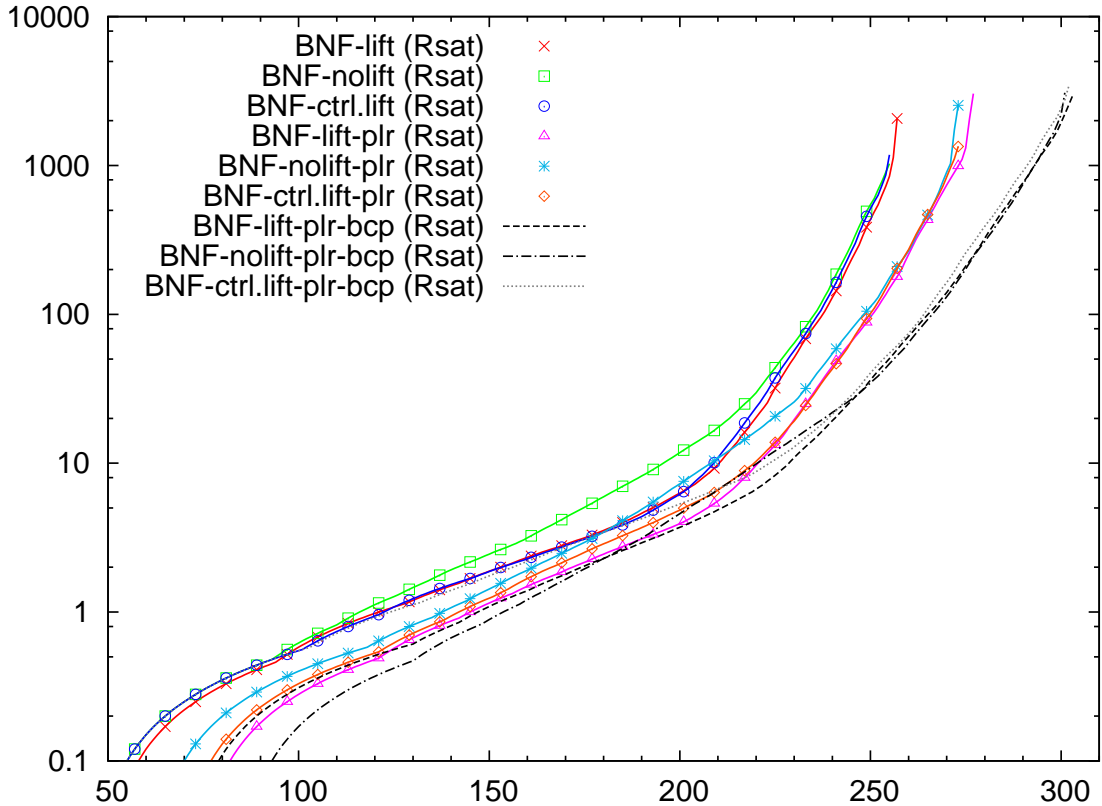

Figure 3: Comparison of different variants of $K_m2SAT$ +Rsat on the LWB problems. $X$ axis: number of solved problems; $Y$ axis: total CPU time spent (sorting problems from the easiest to the hardest).

In general, we do not have relevant performance gaps between the various configurations on this benchmark suite; it seems that in the majority of cases `ctrl.lift` slightly beats `nolift` and `nolift` slightly beats `lift`. These gaps are even more relevant if we consider the size of the formulas generated (Figure 5). We believe that this may be due to the fact that random $\Box_m$-CNF formulas may contain lots of shared subformulas $\Box_r\psi$, so that lifting may cause a reduction of such sharing (see Section 3). Further, `plr-bcp` does not seem to introduce relevant improvements here. We believe that this is due to the fact that these random formulas produce pure and unit literals with very low or even zero probability.

Overall, the configuration `BNF-nolift` turns out to be the best performer on this suite, with a slight advantage wrt. `BNF-ctrl.lift-plr-bcp`.

Finally, from some plots of Figure 4 we notice that for $K_m2SAT$ +Rsat the problems tend to be harder within the satisfiability/unsatisfiability transition area. (This fact holds especially for Racer and *SAT, see Section 5.3.) This seems to confirm the fact that the easy-hard-easy pattern of random k-SAT extends also to $\Box_m$-CNF, as already observed in literature (Giunchiglia & Sebastiani, 1996, 2000; Giunchiglia et al., 2000; Horrocks et al., 2000; Patel-Schneider & Sebastiani, 2003).

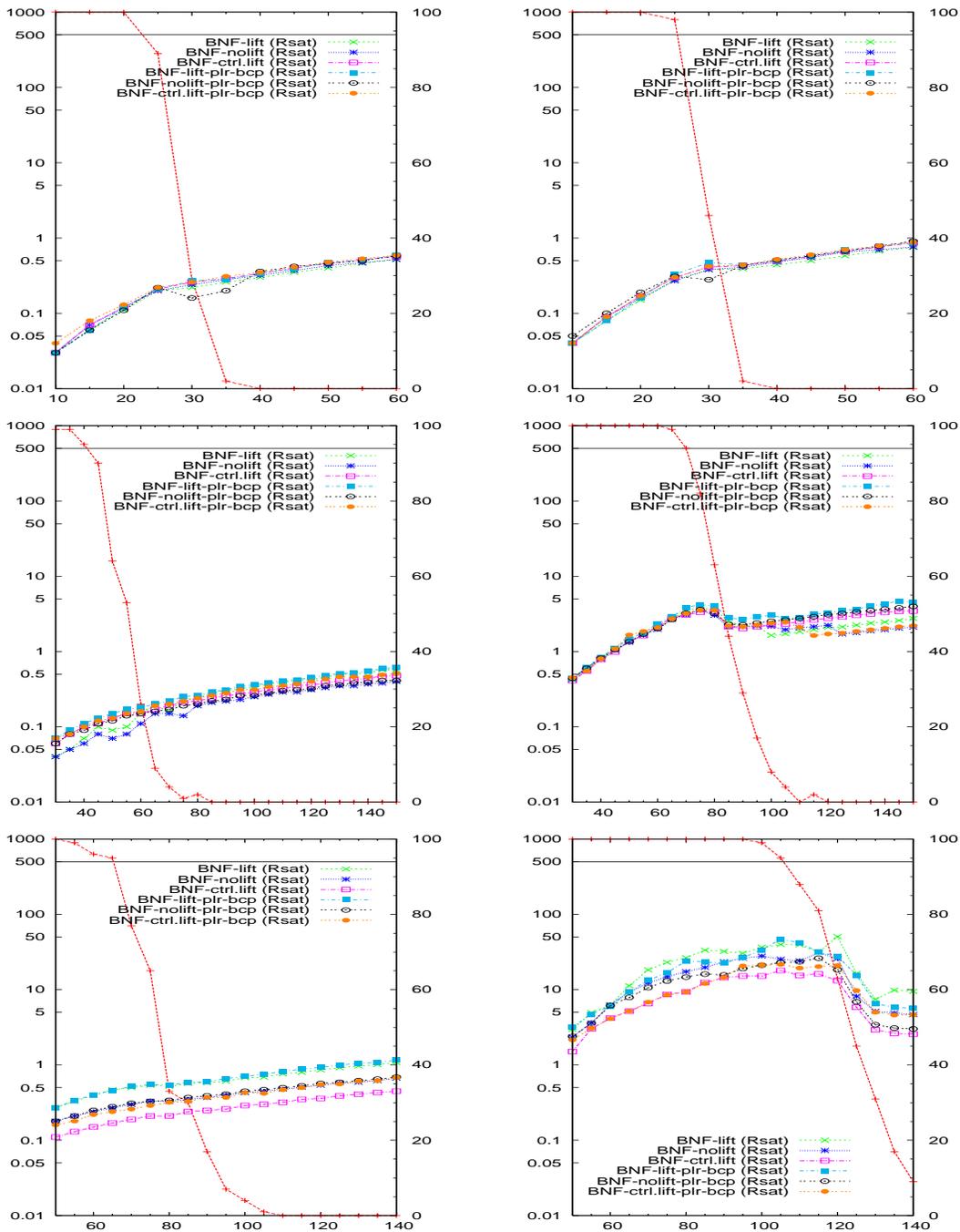

Figure 4: Comparison among different variants of $K_m2SAT$ +RSAT on random problems. X axis: $\#clauses/N$. Y axis: median (50th percentile) CPU time, 100 samples/point. 1st row: $d = 1$, $p = 0.5$, $N = 8, 9$; 2nd row: $d = 2$, $p = 0.6$, $N = 3, 4$; 3rd row: $d = 2$, $p = 0.5$, $N = 3, 4$. Background: % of satisfiable instances.

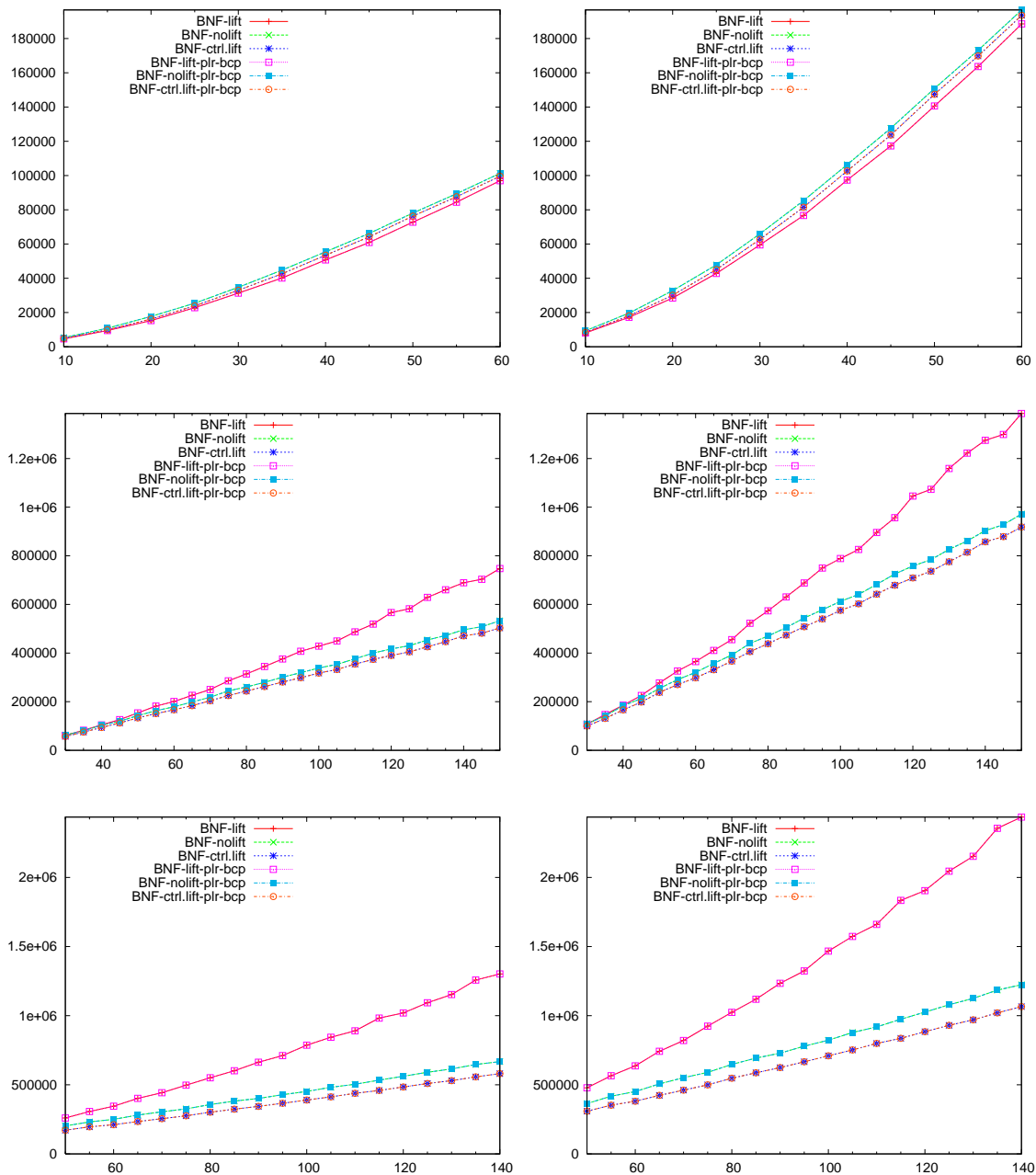

Figure 5: Comparison among different variants of $K_m2SAT$ on random problems. X axis: $\#clauses/N$. Y axis: 1st column: $\#variables$ in the SAT encoding (90th percentiles), 100 samples/point; 2nd column: $\#clauses$ in the SAT encoding (90th percentiles), 100 samples/point. 1st row: $d = 1$, $p = 0.5$, $N = 9$; 2nd row: $d = 2$, $p = 0.6$, $N = 4$; 3rd row: $d = 2$, $p = 0.5$, $N = 4$.

### 5.2.3 Results on the TANCS 2000 Benchmark Suite

The comparison among the $K_m 2SAT$ variants on the TANCS 2000 benchmark is presented in Figures 6 and 7, where different `BNF` variants of $K_m 2SAT$ are compared both enabling or disabling `lift/ctrl.lif` and `plr-bcp`.

In Figure 6, from top-left to bottom, we present the cumulative CPU times spent by $K_m 2SAT$ +Rsat on the easy, medium and hard categories respectively (the corresponding plots reporting the non-cumulative CPU times are included in the online appendix). In Figure 7 we present the plots of the numbers of variables and clauses of the formulas solved for the more challenging cases of the medium and hard problems.[31] We notice that there are only slight differences among the different variants of $K_m 2SAT$; `BNF` with `lift` is the best option which allows for solving more problems in the hard class and requiring less CPU time.

We remark that, despite the expected exponential growth of the encoded formulas wrt. the modal depth, in this benchmark $K_m 2SAT$ +Rsat allows for encoding and solving problems of modal depth greater than 100 for the easy class and greater than 50 for the medium and hard classes, producing and solving SAT-encoded formulas with more than $10^7$ variables and $1.4 \cdot 10^7$ clauses.

## 5.3 An Empirical Comparison wrt. the Other Approaches

We proceed with the comparison of our approach wrt. the current state-of-the-art evaluating $K_m 2SAT$ +Rsat against the other $K_m$-satisfiability solvers listed above. In more details, we chose to compare the performance of the other solvers against both the best "local" $K_m 2SAT$ +Rsat variant for the single benchmark suite and the best "global" $K_m 2SAT$ +Rsat variant among all the benchmark suites, which we have identified in `BNF-ctrl.lift-plr-bcp`.

### 5.3.1 Comparison on the LWB Benchmark Suite

The results on the LWB benchmark suite are summarized numerically and graphically in Table 2. From Table 2(a) we notice a few facts: Racer and *SAT are the best performers (confirming the analysis done by Horrocks et al., 2000) with a significant gap wrt. the others; then, K-QBF +sKizzo solves a few more problems than $K_m 2SAT$ +Rsat; then follows KBDD which outperforms Mspass, which is the worst performer. In detail, $K_m 2SAT$ +Rsat is (one of) the worst performer(s) on `k_d4_*` and `k_t4_*`, the fourth best performer on `k_path_n`, the third best performer on `k_path_p` and `k_branch_p`, and it is (one of) the best performer(s) on `k_branch_n`, `k_dum_*`, `k_grz_*`, `k_lin_*`, `k_ph_*` and `k_poly_*`; it is the absolute best performer on `k_branch_n` and `k_ph_p`.

In Table 2(b) we give a graphical representation of this comparison, plotting the number of solved problems by each approach against the total cumulative amount of CPU time spent. We notice that, even if $K_m 2SAT$ +Rsat solves a few problems less than K-QBF +sKizzo, $K_m 2SAT$ +Rsat is mostly faster than K-QBF +sKizzo.

---

31. The same plots for the easy problems are included in the online appendix.

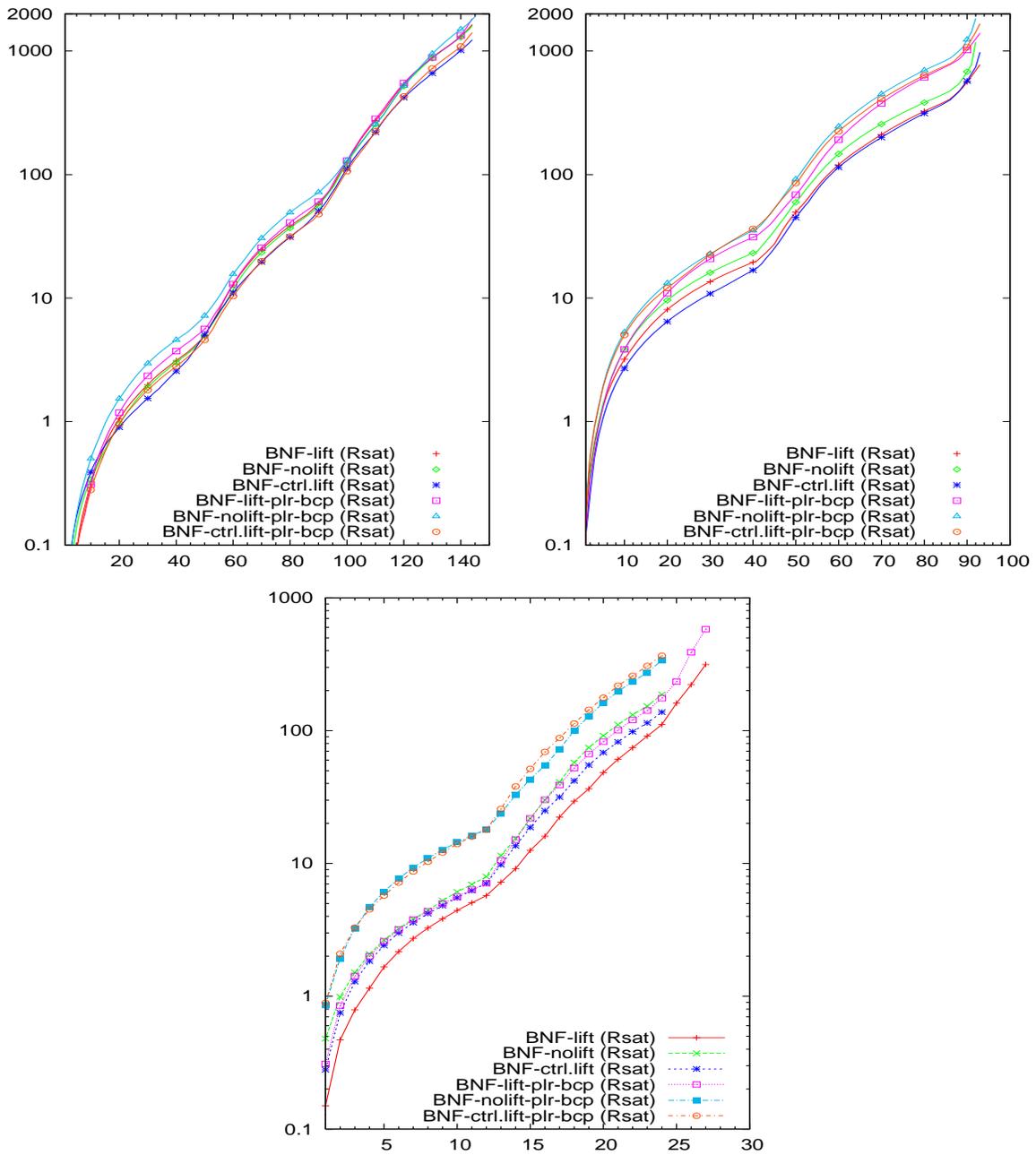

Figure 6: Comparison among different variants of $K_m 2SAT$ +Rsat on TANCS 2000 problems. $X$ axis: number of solved problems. $Y$ axis: total cumulative CPU time spent. 1st (top-left) to 3th (bottom) plot: easy, medium, hard problems. (Problems are sorted from the easiest to the hardest).

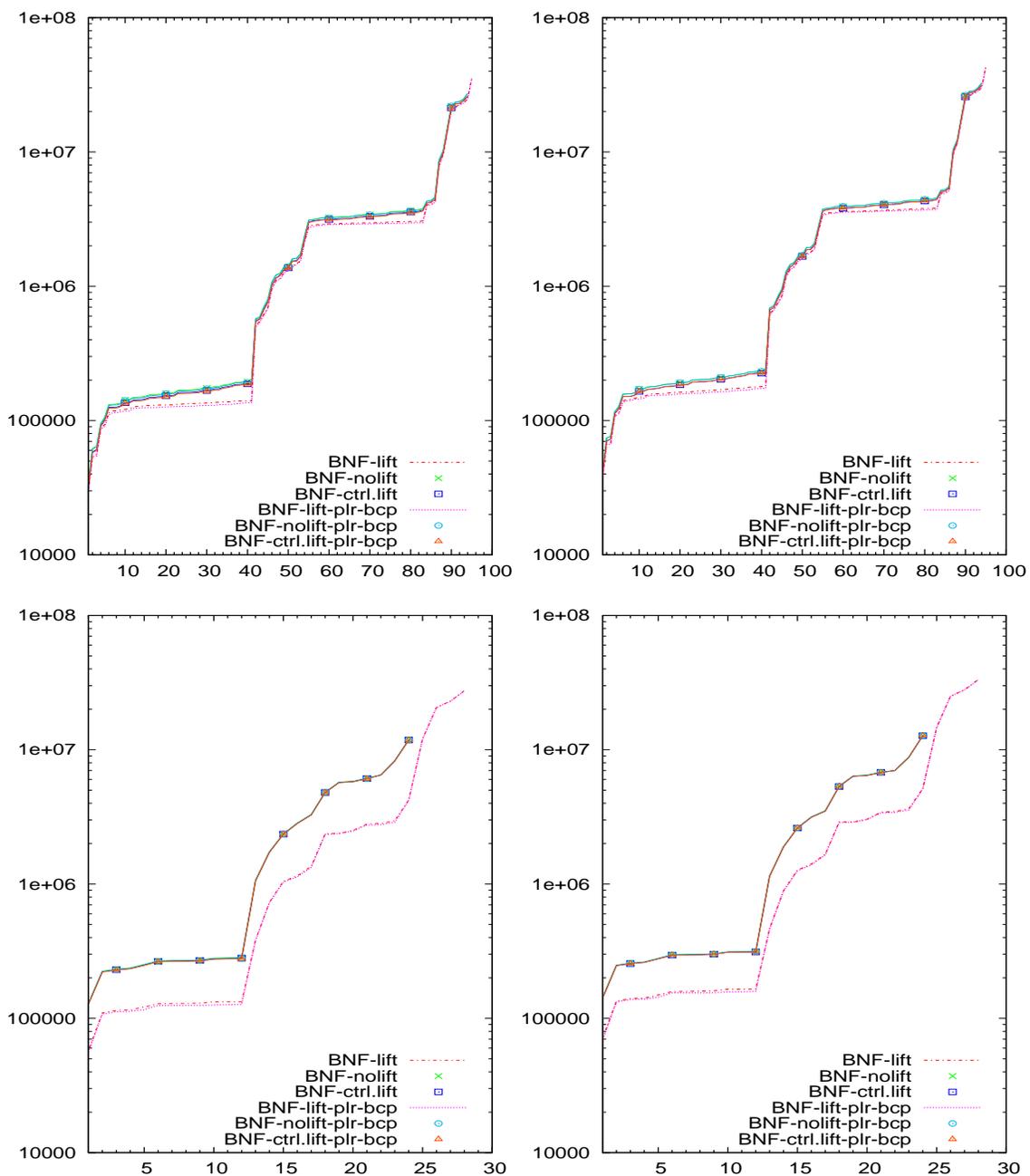

Figure 7: Comparison among different variants of $K_m2SAT$ on TANCS 2000 problems. $X$ axis: number of the harder solved problem. $Y$ axis: 1st column: $\#variables$ in the SAT encoding of the problem; 2nd column: $\#clauses$ in the SAT encoding of the problem. 1st to 2th row: medium, hard problems.

| | other tools | | | | | $K_m2SAT$ + Rsat | |
|---|---|---|---|---|---|---|---|
| | K-QBF | | | | | BNF-plr-bcp | |
| test | + sKizzo | KBDD | Mspass | Racer | *SAT | -ctrl.lift | -lift |
| k_branch_n | 4 | 8 | 10 | 15 | 14 | **17** | **17** |
| k_branch_p | 16 | 8 | 10 | **21** | **21** | 18 | 18 |
| k_d4_n | 8 | **21** | **21** | **21** | **21** | 8 | 8 |
| k_d4_p | **21** | **21** | **21** | **21** | **21** | 14 | 14 |
| k_dum_n | **21** | **21** | **21** | **21** | **21** | **21** | **21** |
| k_dum_p | **21** | **21** | **21** | **21** | **21** | **21** | **21** |
| k_grz_n | 19 | **21** | **21** | **21** | **21** | **21** | **21** |
| k_grz_p | **21** | **21** | **21** | **21** | **21** | **21** | **21** |
| k_lin_n | 20 | **21** | **21** | **21** | **21** | **21** | **21** |
| k_lin_p | **21** | **21** | 3 | **21** | **21** | **21** | **21** |
| k_path_n | 9 | **21** | 4 | **21** | **21** | 13 | 14 |
| k_path_p | 13 | 17 | 5 | **21** | **21** | 15 | 16 |
| k_ph_n | **21** | 4 | 12 | **21** | 13 | **21** | **21** |
| k_ph_p | 10 | 4 | 8 | 9 | 9 | **11** | 10 |
| k_poly_n | **21** | 8 | 7 | **21** | **21** | **21** | **21** |
| k_poly_p | **21** | 8 | 7 | **21** | **21** | **21** | **21** |
| k_t4p_n | **21** | **21** | 12 | **21** | **21** | 6 | 6 |
| k_t4p_p | **21** | **21** | **21** | **21** | **21** | 11 | 11 |

(a) Indexes of the hardest problems solved.

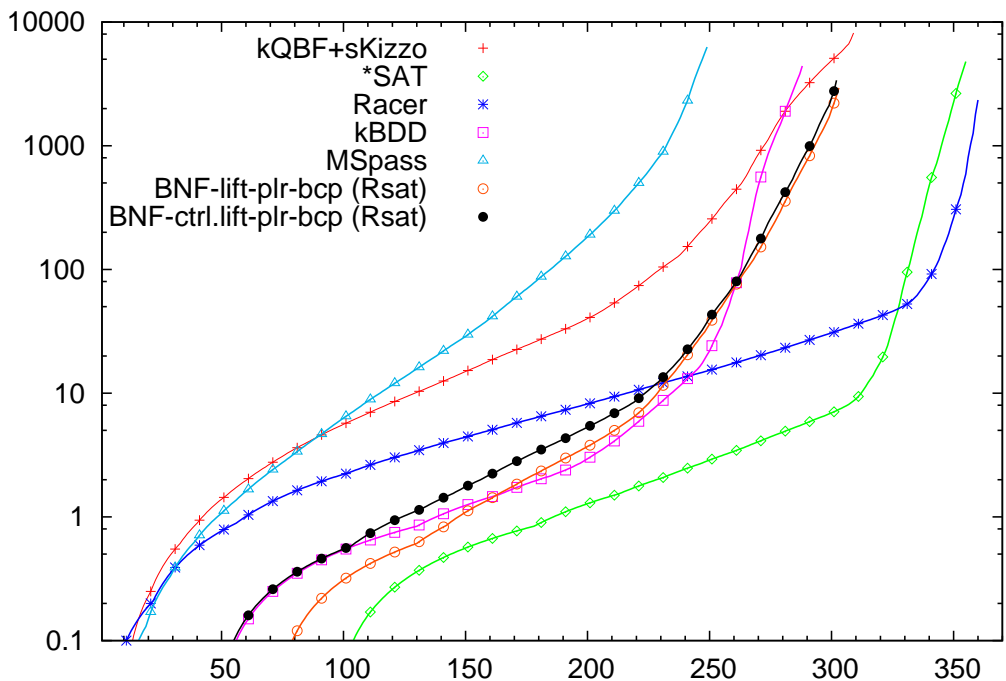

(b) $X$ axis: # of problems solved; $Y$ axis: total (cumulative) CPU time spent.

Table 2: Comparison of $K_m2SAT$ +Rsat against the other tools on the LWB benchmarks.

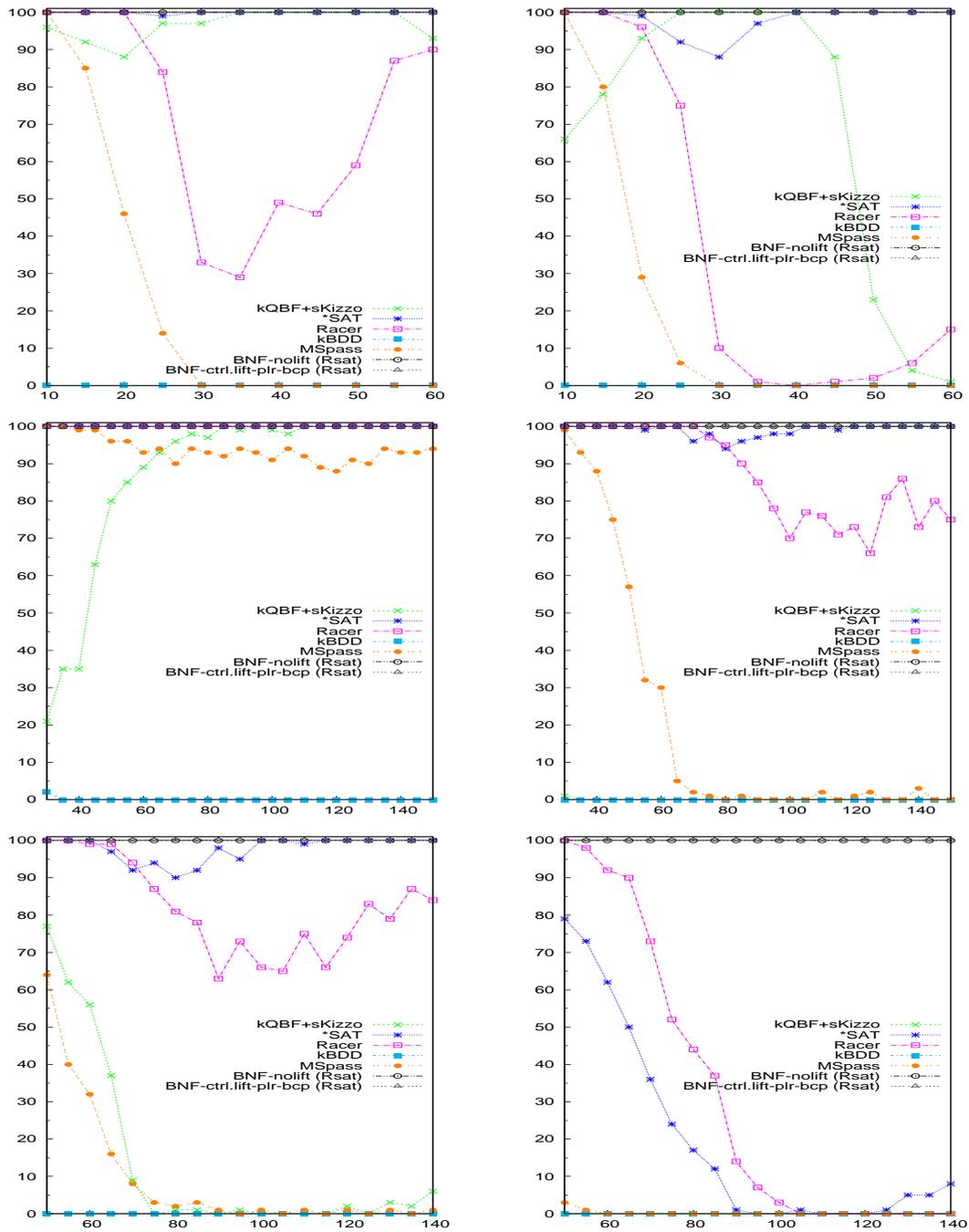

Figure 8: Comparison against other approaches on random problems. X axis: $\#clauses/N$. Y axis: % of problems solved within the timeout, 100 samples/point. 1st row: $d = 1$, $p = 0.5$, $N = 8, 9$; 2nd row: $d = 2$, $p = 0.6$, $N = 3, 4$; 3rd row: $d = 2$, $p = 0.5$, $N = 3, 4$.

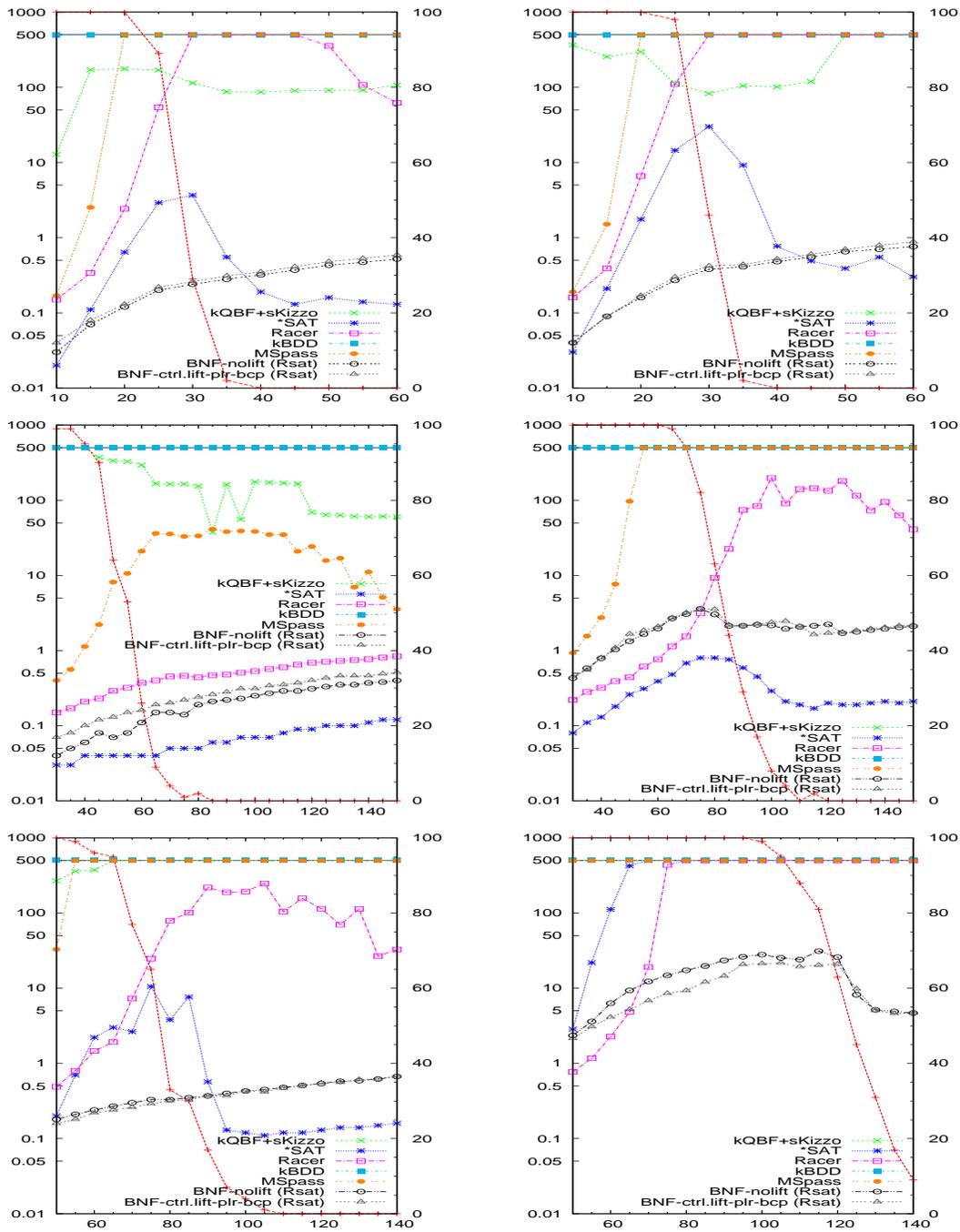

Figure 9: Comparison against other approaches on random problems. X axis: $\#clauses/N$. Y axis: median (50th percentile) CPU time, 100 samples/point. 1st row: $d = 1$, $p = 0.5$, $N = 8, 9$; 2nd row: $d = 2$, $p = 0.6$, $N = 3, 4$; 3rd row: $d = 2$, $p = 0.5$, $N = 3, 4$. Background: % of satisfiable instances.

### 5.3.2 Comparison on the Random $\Box_m$-CNF Benchmark Suite

In the random $\Box_m$-CNF benchmark suite the results are dominated by $K_m2SAT$ +Rsat. For the hardest categories among the three groups of problems used in the evaluation, we report in Figure 8 the number of problems solved by each tool within the timeout, and in Figure 9 the median CPU time (i.e. the 50%th percentile).

Looking at Figure 8 we notice that $K_m2SAT$ +Rsat (in both versions) is the only tool which always terminates within the timeout, whilst *SAT and Racer sometimes do not terminate in the hardest problems, K-QBF +sKizzo very often does not terminate, and Mspass and KBDD do not terminate for most values.

In Figure 9 we notice that $K_m2SAT$ +Rsat is most often the best performer (in particular with the hardest problems), followed by *SAT and Racer. (This is even much more evident if we consider the 90%th percentile of CPU time, whose plots are included in the online appendix.) In all these tests, K-QBF +sKizzo, Mspass and KBDD are drastically outperformed by the others.

### 5.3.3 Comparison on the TANCS 2000 Benchmark Suite

The results of the TANCS 2000 benchmark are summarized in Figure 10, from the easy category (top-left) to the hard category (bottom).

From the plots we notice that the relative performances of the tools under test vary significantly with the category: Racer and *SAT are among the best performers in all categories; K-QBF +sKizzo behaves well on the easy and medium categories but solves very few problems on the hard one; KBDD behaves very well on the easy category, but solves very few problems in the medium and hard ones. Mspass is among the worst performers in all categories: in particular, it does not solve any problem in the hard category; finally, $K_m2SAT$ +Rsat is the worst performer on the easy category, it outperforms all competitors but *SAT and Racer on the medium category, and competes head-to-head with both Racer and *SAT for the first position on the hard category: the "local-best" configuration `BNF-lift` beats both competitors; the "global-best" configuration `BNF-ctrl.lift-prl-bcp` solves as many problems as Racer, with one-order-magnitude CPU-time performance gap, and two problems less than *SAT.

Notice that the classification of the benchmark problems in "easy", "medium" and "hard" given by Massacci and Donini (2000) is based on the translation schema used to produce every modal problem and refers to its "reasoning component", but it is not necessarily related to other components (like, e.g., the modal depth) which affect the size of our encoding and, hence, the efficiency of our approach. This may explain the fact, e.g., that the "easy" problems are not so easy for our approach, and viceversa.

### 5.4 Discussion

As highlighted by Giunchiglia et al. (2000), and Horrocks et al. (2000), the satisfiability problem in modal logics like $K_m$ is characterized by the alternation of two orthogonal components of reasoning: a *Boolean* component, performing Boolean reasoning within each state, and a *modal* component, generating the successor states of each state. The latter must cope with the fact that the candidate models may be up to exponentially big wrt. $depth(\varphi)$, whilst the former must cope with the fact that there may be up to exponentially many candidate (sub)models to explore. In the $K_m2SAT$ +DPLL approach the encoder has to handle the whole modal component (rules (8) and (9)), whilst the handling of the whole Boolean component is delegated to an external SAT solver.

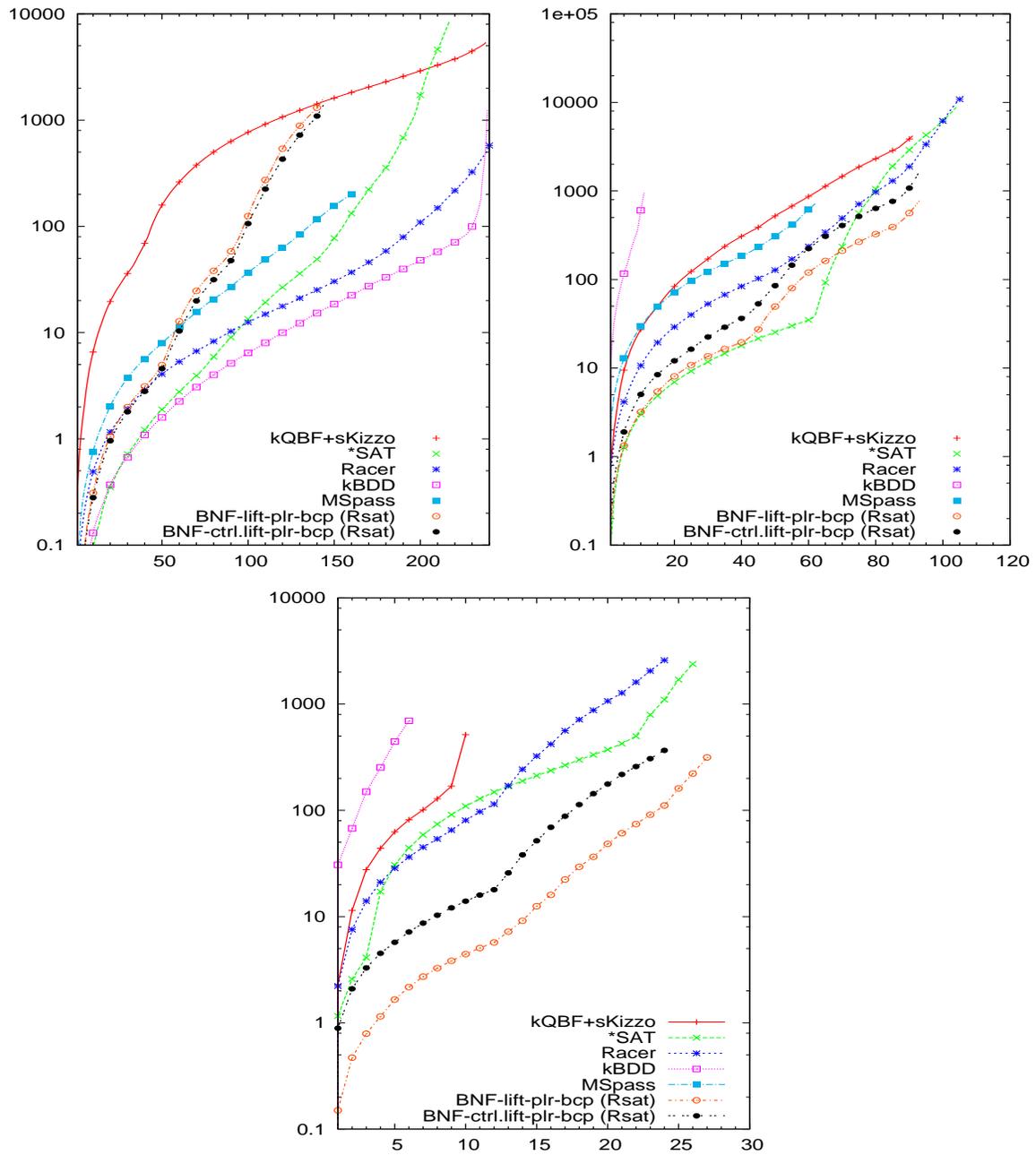

Figure 10: Comparison against other approaches on TANCS 2000 problems. X axis: number of solved problems. Y axis: total cumulative CPU time spent. 1st (top-left) to 3th (bottom) plot: easy, medium, hard problems. (Problems are sorted from the easiest to the hardest).

From the results displayed in Section 5.3 we notice that the relative performances of the $K_m2SAT$ +DPLL approach wrt. other state-of-the-art tools range from cases where $K_m2SAT$ +RSAT is much less efficient than other state-of-the-art approaches (e.g., the k_d4 and k_t4p formulas) up to formulas where it is much more efficient (e.g., the k_ph_p or the $\Box_m$-CNF formulas with $d = 1$). In the middle stands a large majority of formulas in which the approach competes well against the other state-of-the art tools; in particular, $K_m2SAT$ +RSAT competes very well or even outperforms the other approaches based on translations into different formalisms (the translational approach, the automata-theoretic approach and the QBF-encoding approach).

A simple explanation of the former fact could be that the $K_m2SAT$ +DPLL approach loses on problems with high modal depth, or where the modal component of reasoning dominates (like, e.g., the k_d4 and k_t4p formulas), and wins on problems where the Boolean component of reasoning dominates (like, e.g., the k_ph_n or the $\Box_m$-CNF formulas with $d = 1$), and it is competitive for formulas in which both components are relevant.

We notice, however, that $K_m2SAT$ +RSAT wins in the hard category of TANCS 2000 benchmarks, with modal depths greater than 50, and on k_branch_n, where the search is dominated by the modal component. [32] In fact, we recall that reducing the Boolean component of reasoning may produce a reduction also of the modal reasoning effort, because it may reduce the number of successor states to analyze (e.g. Sebastiani, 2007, 2007). Thus, e.g., techniques like on-the-fly BCP, although exploiting only purely-Boolean properties, may produce not only a drastic pruning of the Boolean search, but also a drastic reduction in the size of the model investigated, because they cut a priori the amount of successor states to expand.

## 6. Related Work and Research Trends

In the last fifteen years one main research line in description logic has focused on investigating increasingly expressive logics, with the goal of establishing the theoretical boundaries of decidability and of allowing for more expressive power in the languages defined (Baader, Calvanese, McGuinness, Nardi, & Patel-Schneider, 2003). Consequently, very expressive — though very hard— description logics have today notable application in the field of Semantic Web. For example, the $\mathcal{SHOIN}^{(\mathcal{D})}$ logic (which has NEXPTIME complexity) captures the sub-language OWL DL of the full OWL (Web Ontology Language) language (Bechhofer, van Harmelen, Hendler, Horrocks, McGuinness, Patel-Schneider, & Stein, 2004), that is the recommended standard language for the semantic web proposed by the W3C consortium.

In contrast, the recent quest for tractable description logic-based languages arising from the field of bio-medical ontologies (e.g., Spackman, Campbell, & Cote, 1997; Sioutos, de Coronado, Haber, Hartel, Shaiu, & Wright, 2007; The Gene Ontology Consortium, 2000;

---

32. The k_branch_n formulas are very hard from the perspective of modal reasoning, because they require finding one model $\mathcal{M}$ with $2^{d+1}-1$ states (Halpern & Moses, 1992), but no Boolean reasoning within each state is really required (Giunchiglia et al., 2000; Horrocks et al., 2000): e.g., *SAT solves k_branch_n(d) with $2^{d+1}-1$ calls to its embedded DPLL engine, one for each state of $\mathcal{M}$, each call solved by BCP only.

Rector & Horrocks, 1997) has stimulated the opening of another research line on tractable description logics (also called *lightweight description logics*), which are suitable for reasoning on these very big bio-medical ontologies. In particular, Baader et al. (2005, 2006, 2007, 2008) have spent a considerable effort in the attempt of defining a small but maximal subset of logical constructors, expressive enough to cover the needs of these practical applications, but whose inference problems must remain tractable. For example, simple and tractable description logics like $\mathcal{EL}$, $\mathcal{EL}^+$ and $\mathcal{EL}^{++}$ (Baader et al., 2005) are expressive enough to describe several important bio-medical ontologies such as SNoMed (Spackman et al., 1997), NCI (Sioutos et al., 2007), the Gene Ontology (The Gene Ontology Consortium, 2000) and the majority of Galen (Rector & Horrocks, 1997).

Reasoning on these ontologies represents not only an important application of lightweight description logics, but also a challenge due to the required efficiency and the huge dimensions of the ontologies. In this perspective, it is important to face efficiently not only the basic reasoning services (e.g., satisfiability, subsumption, queries) on logics like $\mathcal{EL}$, $\mathcal{EL}^+$ and $\mathcal{EL}^{++}$, but also more sophisticated reasoning problems like e.g., *axiom pinpointing* (Baader et al., 2007; Baader & Peñaloza, 2008) and *logical difference* between terminologies (Konev, Walther, & Wolter, 2008).

## 7. Conclusions and Future Work

In this paper we have explored the idea of encoding $K_m/\mathcal{ALC}$-satisfiability into SAT, so that to be handled by state-of-the-art SAT tools. We have showed that, despite the intrinsic risk of blowup in the size of the encoded formulas, the performances of this approach are comparable with those of current state-of-the-art tools on a rather extensive variety of empirical tests. Furthermore, we remark that our approach allows for directly benefitting "for free" from current and future enhancements in SAT solver technology.

We see many possible directions to explore in order to enhance and extend our approach. An important open research line is to explore the feasibility of SAT encodings for other and more expressive modal and description logics (e.g., whilst for logics like $T_m$ the extension should be straightforward, logics like $S4_m$, or more elaborated description logics than $\mathcal{ALC}$, should be challenging) and for more complex form of reasoning (e.g., including TBoxes and ABoxes).

Our current investigation, however, focuses on the lightweight logics of Baader et al. (2005). We have investigated (and we are currently enhancing) an encoding of the main reasoning services in $\mathcal{EL}$ and $\mathcal{EL}^+$ into Horn-SAT, which allows for reasoning efficiently on the (often huge) bio-medical ontologies mentioned in Section 6, and for handling the more sophisticated inference problems mentioned there (e.g., we currently handle axiom pinpointing), by exploiting some of the advanced functionalities which can be implemented on top of a modern SAT solver (Sebastiani & Vescovi, 2009).

## 8. Acknowledgments

The authors are partly supported by SRC/GRC under Custom Research Project 2009-TJ-1880 WOLFLING, and by MIUR under PRIN project 20079E5KM8_002.

# Appendix A. The Proof of Correctness & Completeness

## A.1 Some Further Notation

Let $\psi$ be a $K_m$-formula. We denote by $\overline{\psi}$ the representation of $\neg\psi$ in the current formalism: in NNF, $\overline{\Diamond_r\psi} \stackrel{\text{def}}{=} \Box_r\overline{\psi}$, $\overline{\Box_r\psi} \stackrel{\text{def}}{=} \Diamond_r\overline{\psi}$, $\overline{\psi_1 \wedge \psi_2} \stackrel{\text{def}}{=} \overline{\psi_1} \vee \overline{\psi_2}$, $\overline{\psi_1 \vee \psi_2} \stackrel{\text{def}}{=} \overline{\psi_1} \wedge \overline{\psi_2}$, $\overline{A_i} \stackrel{\text{def}}{=} \neg A_i$, $\overline{\neg A_i} \stackrel{\text{def}}{=} A_i$; in BNF, $\overline{\neg\Box_r\psi} \stackrel{\text{def}}{=} \Box_r\psi$, $\overline{\Box_r\psi} \stackrel{\text{def}}{=} \neg\Box_r\psi$, $\overline{\psi_1 \wedge \psi_2} \stackrel{\text{def}}{=} \overline{\psi_1} \vee \overline{\psi_2}$, $\overline{\psi_1 \vee \psi_2} \stackrel{\text{def}}{=} \overline{\psi_1} \wedge \overline{\psi_2}$, $\overline{A_i} \stackrel{\text{def}}{=} \neg A_i$, $\overline{\neg A_i} \stackrel{\text{def}}{=} A_i$.

We represent a truth assignment $\mu$ as a set of literals, with the intended meaning that a positive literal $A_i$ (resp. negative literal $\neg A_i$) in $\mu$ means that $A_i$ is assigned to true (resp. false). We say that $\mu$ *assigns* a literal $l$ if either $l \in \mu$ or $\neg l \in \mu$. We say that a literal $l$ occurs in a Boolean formula $\phi$ iff the atom of $l$ occurs in $\phi$.

Let $\mathcal{M}$ denote a Kripke model, and let $\sigma$ be the label of a generic state $u_\sigma$ in $\mathcal{M}$. We label (and denote) by "1" the root state of $\mathcal{M}$. By "$\langle\sigma:\psi\rangle \in \mathcal{M}$" we mean that $u_\sigma \in \mathcal{M}$ and $\mathcal{M}, u_\sigma \models \psi$. Thus, for every $\sigma$ s.t. $u_\sigma \in \mathcal{M}$, either $\langle\sigma:\psi\rangle \in \mathcal{M}$ or $\langle\sigma:\overline{\psi}\rangle \in \mathcal{M}$.

For convenience, instead of (9) sometimes we use the equivalent definition:

$$Def(\sigma,\ \nu^r) \stackrel{\text{def}}{=} (L_{\langle\sigma,\ \nu^r\rangle} \rightarrow \bigwedge_{\substack{\text{for every} \\ \langle\sigma,\pi^{r,i}\rangle}} (L_{\langle\sigma,\ \pi^{r,i}\rangle} \rightarrow L_{\langle\sigma.i,\ \nu_0^r\rangle})) \wedge \bigwedge_{\substack{\text{for every} \\ \langle\sigma,\pi^{r,i}\rangle}} Def(\sigma.i,\ \nu_0^r). \quad (20)$$

Notice that each $Def(\sigma,\ \psi)$ in (6), (7), (8), (20) is written in the general form

$$(L_{\langle\sigma,\ \psi\rangle} \rightarrow \Phi_{\langle\sigma,\psi\rangle}) \wedge \bigwedge_{\langle\sigma',\psi'\rangle} Def(\sigma',\ \psi'). \quad (21)$$

We call *definition implication* for $Def(\sigma,\ \psi)$ the expressions "$(L_{\langle\sigma,\ \psi\rangle} \rightarrow \Phi_{\langle\sigma,\psi\rangle})$".

## A.2 Soundness and Completeness of $K_m 2SAT$

Let $\varphi$ be a $K_m$-formula. We prove the following theorem, which states the soundness and completeness of $K_m 2SAT$.

**Theorem 1.** *A $K_m$-formula $\varphi$ is $K_m$-satisfiable if and only if the corresponding $K_m 2SAT(\varphi)$ is satisfiable.*

**Proof.** It is a direct consequence of the following Lemmas 2 and 3. □

**Lemma 2.** *Given a $K_m$-formula $\varphi$, if $K_m 2SAT(\varphi)$ is satisfiable, then there exists a Kripke model $\mathcal{M}$ s.t. $\mathcal{M}, 1 \models \varphi$.*

**Proof.** Let $\mu$ be a total truth assignment satisfying $K_m 2SAT(\varphi)$. We build from $\mu$ a Kripke model $\mathcal{M} = \langle \mathcal{U}, \mathcal{L}, \mathcal{R}_1, \ldots, \mathcal{R}_m \rangle$ as follows:

$$\mathcal{U} \stackrel{\text{def}}{=} \{\sigma\ :\ A_{\langle\sigma,\ \psi\rangle}\ \text{occurs in}\ K_m 2SAT(\varphi)\ \text{for some}\ \psi\} \quad (22)$$

$$\mathcal{L}(\sigma, A_i) \stackrel{\text{def}}{=} \begin{cases} True & \text{if}\ L_{\langle\sigma,\ A_i\rangle} \in \mu \\ False & \text{if}\ \neg L_{\langle\sigma,\ A_i\rangle} \in \mu \end{cases} \quad (23)$$

$$\mathcal{R}_r \stackrel{\text{def}}{=} \{\langle\sigma, \sigma.i\rangle\ :\ L_{\langle\sigma,\ \pi^{r,i}\rangle} \in \mu\}. \quad (24)$$

We show by induction on the structure of $\varphi$ that, for every $\langle \sigma, \psi \rangle$ s.t. $L_{\langle \sigma, \psi \rangle}$ occurs on $K_m 2SAT(\varphi)$,

$$\langle \sigma : \psi \rangle \in \mathcal{M} \quad \text{if} \quad L_{\langle \sigma, \psi \rangle} \in \mu. \tag{25}$$

**Base**

$\psi = A_i$ or $\psi = \neg A_i$. Then (25) follows trivially from (23).

**Step**

$\psi = \alpha$. Let $L_{\langle \sigma, \alpha \rangle} \in \mu$.

    As $\mu$ satisfies (6), $L_{\langle \sigma, \alpha_i \rangle} \in \mu$ for every $i \in \{1, 2\}$.

    By inductive hypothesis, $\langle \sigma : \alpha_i \rangle \in \mathcal{M}$ for every $i \in \{1, 2\}$.

    Then, by definition, $\langle \sigma : \alpha \rangle \in \mathcal{M}$.

    Thus, $\langle \sigma : \alpha \rangle \in \mathcal{M}$ if $L_{\langle \sigma, \alpha \rangle} \in \mu$.

$\psi = \beta$. Let $L_{\langle \sigma, \beta \rangle} \in \mu$.

    As $\mu$ satisfies (7), $L_{\langle \sigma, \beta_i \rangle} \in \mu$ for some $i \in \{1, 2\}$.

    By inductive hypothesis, $\langle \sigma : \beta_i \rangle \in \mathcal{M}$ for some $i \in \{1, 2\}$.

    Then, by definition, $\langle \sigma : \beta \rangle \in \mathcal{M}$.

    Thus, $\langle \sigma : \beta \rangle \in \mathcal{M}$ if $L_{\langle \sigma, \beta \rangle} \in \mu$.

$\psi = \pi^{r,j}$. Let $L_{\langle \sigma, \pi^{r,j} \rangle} \in \mu$.

    As $\mu$ satisfies (8), $L_{\langle \sigma.j, \pi_0^{r,j} \rangle} \in \mu$.

    By inductive hypothesis, $\langle \sigma.j : \pi_0^{r,j} \rangle \in \mathcal{M}$.

    Then, by definition and by (24), $\langle \sigma : \pi^{r,j} \rangle \in \mathcal{M}$.

    Thus, $\langle \sigma : \pi^{r,j} \rangle \in \mathcal{M}$ if $L_{\langle \sigma, \pi^{r,j} \rangle} \in \mu$.

$\psi = \nu^r$. Let $L_{\langle \sigma, \nu^r \rangle} \in \mu$.

    As $\mu$ satisfies (9), for every $\langle \sigma, \pi^{r,i} \rangle$ s.t. $L_{\langle \sigma, \pi^{r,i} \rangle} \in \mu$, we have that $L_{\langle \sigma.i, \nu_0^r \rangle} \in \mu$.

    By inductive hypothesis, we have that $\langle \sigma : \pi^{r,i} \rangle \in \mathcal{M}$ and $\langle \sigma.i : \nu_0^r \rangle \in \mathcal{M}$.

    Then, by definition and by (24), $\langle \sigma : \nu^r \rangle \in \mathcal{M}$.

    Thus, $\langle \sigma : \nu^r \rangle \in \mathcal{M}$ if $L_{\langle \sigma, \nu^r \rangle} \in \mu$.

If $\mu \models K_m 2SAT(\varphi)$, then $A_{\langle 1, \varphi \rangle} \in \mu$. Thus, by (25), $\langle 1 : \varphi \rangle \in \mathcal{M}$, i.e., $\mathcal{M}, 1 \models \varphi$. $\qquad \square$

**Lemma 3.** *Given a $K_m$-formula $\varphi$, if there exists a Kripke model $\mathcal{M}$ s.t. $\mathcal{M}, 1 \models \varphi$, then $K_m2SAT(\varphi)$ is satisfiable.*

*Proof.* Let $\mathcal{M}$ be a Kripke model s.t. $\mathcal{M}, 1 \models \varphi$. We build from $\mathcal{M}$ a truth assignment $\mu$ for $K_m2SAT(\varphi)$ recursively as follows: [33]

$$\mu \stackrel{\text{def}}{=} \mu_{\mathcal{M}} \cup \mu_{\overline{\mathcal{M}}} \tag{26}$$

$$\mu_{\mathcal{M}} \stackrel{\text{def}}{=} \{L_{\langle \sigma, \ \psi \rangle} \in K_m2SAT(\varphi) \ : \ \langle \sigma, \psi \rangle \in \mathcal{M}\} \tag{27}$$
$$\cup \ \{\neg L_{\langle \sigma, \ \psi \rangle} \in K_m2SAT(\varphi) \ : \ \langle \sigma, \overline{\psi} \rangle \in \mathcal{M}\}$$

$$\mu_{\overline{\mathcal{M}}} \stackrel{\text{def}}{=} \mu_{\pi\nu} \cup \mu_{\alpha\beta} \cup \mu_{\mathcal{A}} \tag{28}$$

$$\mu_{\pi\nu} \stackrel{\text{def}}{=} \{\neg L_{\langle \sigma, \ \pi^{r,i} \rangle} \in K_m2SAT(\varphi) \ : \ \sigma \notin \mathcal{M}\} \tag{29}$$
$$\cup \ \{L_{\langle \sigma, \ \nu^r \rangle} \in K_m2SAT(\varphi) \ : \ \sigma \notin \mathcal{M}\}$$

$$\mu_{\alpha\beta} \stackrel{\text{def}}{=} \{\neg L_{\langle \sigma, \ \alpha \rangle} \in K_m2SAT(\varphi) \ : \ \sigma \notin \mathcal{M} \text{ and } \neg L_{\langle \sigma, \ \alpha_i \rangle} \in \mu_{\overline{\mathcal{M}}} \text{ for some } i \in \{1,2\}\} \tag{30}$$
$$\cup \ \{\neg L_{\langle \sigma, \ \beta \rangle} \in K_m2SAT(\varphi) \ : \ \sigma \notin \mathcal{M} \text{ and } \neg L_{\langle \sigma, \ \beta_i \rangle} \in \mu_{\overline{\mathcal{M}}} \text{ for every } i \in \{1,2\}\}.$$

where $\mu_{\mathcal{A}}$ is a consistent truth assignment for the literals $L_{\langle \sigma, \ A_i \rangle}$ s.t. $A_i \in \mathcal{A}$ and $\sigma \notin \mathcal{M}$.

By construction, for every $L_{\langle \sigma, \ \psi \rangle}$ in $K_m2SAT(\varphi)$, $\mu$ assigns $L_{\langle \sigma, \ \psi \rangle}$ to true iff it assigns $L_{\langle \sigma, \ \overline{\psi} \rangle}$ to false and vice versa, so that $\mu$ is a consistent truth assignment.

First, we show that $\mu_{\mathcal{M}}$ satisfies the definition implications of all $Def(\sigma, \ \psi)$'s and $Def(\sigma, \ \overline{\psi})$' s.t. $\sigma \in \mathcal{M}$. Let $\sigma \in \mathcal{M}$. We distinguish four cases.

$\psi = \alpha$. Thus $\overline{\psi} = \beta$ s.t. $\beta_1 = \overline{\alpha_1}$ and $\beta_2 = \overline{\alpha_2}$.

- If $\langle \sigma : \alpha \rangle \in \mathcal{M}$ (and hence $\langle \sigma : \beta \rangle \notin \mathcal{M}$), then for both $i$'s $\langle \sigma : \alpha_i \rangle \in \mathcal{M}$ and $\langle \sigma : \beta_i \rangle \notin \mathcal{M}$. Thus, by (27), $\{L_{\langle \sigma, \ \alpha_1 \rangle}, L_{\langle \sigma, \ \alpha_2 \rangle}, \neg L_{\langle \sigma, \ \beta \rangle}\} \subseteq \mu_{\mathcal{M}}$, so that $\mu_{\mathcal{M}}$ satisfies the definition implications of both $Def(\sigma, \ \alpha)$ and $Def(\sigma, \ \beta)$.

- If $\langle \sigma : \alpha \rangle \notin \mathcal{M}$ (and hence $\langle \sigma, \beta \rangle \in \mathcal{M}$), then for some $i$ $\langle \sigma : \alpha_i \rangle \notin \mathcal{M}$ and $\langle \sigma : \beta_i \rangle \in \mathcal{M}$. Thus, by (27), $\{\neg L_{\langle \sigma, \ \alpha \rangle}, L_{\langle \sigma, \ \beta_i \rangle}\} \subseteq \mu_{\mathcal{M}}$, so that $\mu_{\mathcal{M}}$ satisfies the definition implications of both $Def(\sigma, \ \alpha)$ and $Def(\sigma, \ \beta)$.

$\psi = \beta$. Like in the previous case, inverting $\psi$ and $\overline{\psi}$.

$\psi = \pi^{r,j}$. Thus $\overline{\psi} = \nu^r$ s.t. $\nu_0^r = \overline{\pi_0^{r,j}}$.

- If $\langle \sigma : \pi^{r,j} \rangle \in \mathcal{M}$ (and hence $\langle \sigma : \nu^r \rangle \notin \mathcal{M}$), then $\langle \sigma.j : \pi_0^{r,j} \rangle \in \mathcal{M}$. Thus, by (27), $\{L_{\langle \sigma.j, \ \pi_0^{r,j} \rangle}, \neg L_{\langle \sigma, \ \nu^r \rangle}\} \subseteq \mu_{\mathcal{M}}$, so that $\mu_{\mathcal{M}}$ satisfies the definition implications of both $Def(\sigma, \ \pi^{r,j})$ and $Def(\sigma, \ \nu^r)$.

---

33. We assume that $\mu_{\mathcal{M}}$, $\mu_{\pi\nu}$ and $\mu_{\alpha\beta}$ are generated *in order*, so that $\mu_{\alpha\beta}$ is generated recursively starting from $\mu_{\pi\nu}$. Intuitively, $\mu_{\mathcal{M}}$ assigns the literals $L_{\langle \sigma, \ \psi \rangle}$ s.t. $\sigma \in \mathcal{M}$ in such a way to mimic $\mathcal{M}$; $\mu_{\overline{\mathcal{M}}}$ assigns the other literals in such a way to mimic the fact that no state outside those in $\mathcal{M}$ is generated (i.e., all $L_{\langle \sigma, \ \pi \rangle}$'s are assigned false and the $L_{\langle \sigma, \ \nu \rangle}$'s, $L_{\langle \sigma, \ \alpha \rangle}$'s, $L_{\langle \sigma, \ \beta \rangle}$'s are assigned consequently).

– If $\langle \sigma : \pi^{r,j} \rangle \notin \mathcal{M}$ (and hence $\langle \sigma : \nu^r \rangle \in \mathcal{M}$), then by (27) $\neg L_{\langle \sigma, \pi^{r,j} \rangle} \in \mu_{\mathcal{M}}$, so that $\mu_{\mathcal{M}}$ satisfies the definition implications of $Def(\sigma, \pi^{r,j})$.

As far as $Def(\sigma, \nu^r)$ is concerned, we partition the clauses in (9):

$$((L_{\langle \sigma, \nu^r \rangle} \wedge L_{\langle \sigma, \pi^{r,i} \rangle}) \rightarrow L_{\langle \sigma.i, \nu_0^r \rangle}) \tag{31}$$

into two subsets. The first is the set of clauses (31) for which $\langle \sigma : \pi^{r,i} \rangle \in \mathcal{M}$. As $\langle \sigma : \nu^r \rangle \in \mathcal{M}$, we have that $\langle \sigma.i : \nu_0^r \rangle \in \mathcal{M}$. Thus, by (27), $L_{\langle \sigma.i, \nu_0^r \rangle} \in \mu_{\mathcal{M}}$, so that $\mu_{\mathcal{M}}$ satisfies (31). The second is the set of clauses (31) for which $\langle \sigma : \pi^{r,i} \rangle \notin \mathcal{M}$. By (27) we have that $\neg L_{\langle \sigma, \pi^{r,i} \rangle} \in \mu_{\mathcal{M}}$, so that $\mu_{\mathcal{M}}$ satisfies (31). Thus, $\mu_{\mathcal{M}}$ satisfies the definition implications also of $Def(\sigma, \nu^r)$.

$\psi = \nu^r$. Like in the previous case, inverting $\psi$ and $\overline{\psi}$.

Notice that, if $\sigma \notin \mathcal{M}$, then $\sigma.i \notin \mathcal{M}$ for every $i$. Thus, for every $Def(\sigma, \psi)$ s.t. $\sigma \notin \mathcal{M}$, all atoms in the implication definition of $Def(\sigma, \psi)$ are not assigned by $\mu_{\mathcal{M}}$.

Second, we show by induction on the recursive structure of $\mu_{\overline{\mathcal{M}}}$ that $\mu_{\overline{\mathcal{M}}}$ satisfies the definition implications of all $Def(\sigma, \psi)$'s and $Def(\sigma, \overline{\psi})$'s s.t. $\sigma \notin \mathcal{M}$. Let $\sigma \notin \mathcal{M}$.

As a base step, by (29), $\mu_{\pi\nu}$ satisfies the definition implications of all $Def(\sigma, \pi^{r,i})$'s and $Def(\sigma, \nu^r)$'s because it assigns false to all $L_{\langle \sigma, \pi^{r,i} \rangle}$'s. Indeed, $\mu_{\mathcal{A}}$ assigns every literal of the type $L_{\langle \sigma, A_i \rangle}$ s.t $A_i \in \mathcal{A}$ and $\sigma \notin \mathcal{M}$ (notice that all the $Def(\sigma, A_i)$'s definitions are trivially satisfied and don't contain any definition implications).

As inductive step, we show on the inductive structure of $\mu_{\alpha\beta}$ that $\mu_{\alpha\beta}$ satisfies the definition implications of all $Def(\sigma, \alpha)$'s and $Def(\sigma, \beta)$'s

Let $\psi \stackrel{\text{def}}{=} \alpha$ and $\overline{\psi} = \beta$ s.t. $\beta_i = \overline{\alpha_i}$ (or vice versa). Then we have that:

- if both $L_{\langle \sigma, \alpha_i \rangle}$'s (respectively at least one $L_{\langle \sigma, \beta_i \rangle}$) are assigned true by $\mu_{\overline{\mathcal{M}}}$, then the definition implications of $Def(\sigma, \alpha)$ (respectively $Def(\sigma, \beta)$) is already trivially satisfied;

- if at least one $L_{\langle \sigma, \alpha_i \rangle}$ (respectively both $L_{\langle \sigma, \beta_i \rangle}$'s) is assigned false by $\mu_{\overline{\mathcal{M}}}$, then by (30) $L_{\langle \sigma, \alpha \rangle}$ (respectively $L_{\langle \sigma, \beta \rangle}$) is assigned false by $\mu_{\alpha\beta}$, which satisfies the definition implication of $Def(\sigma, \alpha)$ (respectively $Def(\sigma, \beta)$).

Thus $\mu_{\overline{\mathcal{M}}}$ satisfies the definition implications of all the $Def(\sigma, \psi)$'s and $Def(\sigma, \overline{\psi})$'s s.t. $\sigma \notin \mathcal{M}$.

On the whole, $\mu \models Def(\sigma, \psi)$ for every $Def(\sigma, \psi)$. By construction, $\mu_{\mathcal{M}} \models A_{\langle 1, \varphi \rangle}$ since $\langle 1 : \varphi \rangle \in \mathcal{M}$. Therefore $\mu \models K_m2SAT(\varphi)$. $\quad\square$